\documentclass[10pt]{iopart}

\usepackage{iopams}  

\expandafter\let\csname equation*\endcsname\relax
\expandafter\let\csname endequation*\endcsname\relax
\usepackage{amsmath}

\usepackage{mathtools}

\usepackage{graphicx}
\usepackage{subfig}
\usepackage{float}
\usepackage{tikz}

\usepackage{booktabs}
\usepackage{multirow}
	
\usepackage[hyphens]{url}
\usepackage[hidelinks]{hyperref}
\hypersetup{breaklinks=true}

\newcommand{\mean}[1]{\langle #1\rangle}
\newcommand{\mb}[1]{\mathbf{#1}}

\newcommand{\bds}[1]{\boldsymbol{#1}}

\newcommand{\f}[2]{\frac{#1}{#2}}
\newcommand{\abs}[1]{\vert #1\vert}

\newcommand{\vekk}[1]{\mathbf{#1}}

\begin{document}

\title{The view of TK-SVM on the phase hierarchy in the classical kagome Heisenberg antiferromagnet}

\author{Jonas Greitemann$^{1,2}$, Ke Liu$^{1,2,*}$, Lode Pollet$^{1,2,3}$}
\address{$^1$ Arnold Sommerfeld Center for Theoretical Physics, University of Munich, Theresienstr. 37, 80333 M\"unchen, Germany}
\address{$^2$ Munich Center for Quantum Science and Technology (MCQST), Schellingstr. 4, 80799 M\"unchen, Germany}
\address{$^3$ Wilczek Quantum Center, School of Physics and Astronomy, Shanghai Jiao Tong University, Shanghai 200240, China}
\ead{ke.liu@lmu.de}

\vspace{10pt}
\begin{indented}
\item[]\today
\end{indented}

\begin{abstract}
We illustrate how the tensorial kernel support vector machine (TK-SVM) can probe the hidden multipolar orders and emergent local constraint in the classical kagome Heisenberg antiferromagnet.
We show that TK-SVM learns the finite-temperature phase diagram in an unsupervised way.
Moreover, in virtue of its strong interpretability, it identifies the tensorial quadrupolar and octupolar orders, which define a biaxial $D_{3h}$ spin nematic, and the local constraint that underlies the selection of coplanar states. We then discuss the disorder hierarchy of the phases, which can be inferred from both the analytical order parameters and a SVM bias parameter.
For completeness we mention that  the machine also picks up the leading $\sqrt{3} \times \sqrt{3}$  correlations in the dipolar channel at very low temperature, which are however weak compared to the quadrupolar and octupolar orders.
Our work shows how TK-SVM can facilitate and speed up the analysis of classical frustrated magnets.
\end{abstract}

%
\ioptwocol

\section{Introduction} \label{sec:intro}
\begin{figure}[t]
  \centering
  \includegraphics[width=0.48\textwidth]{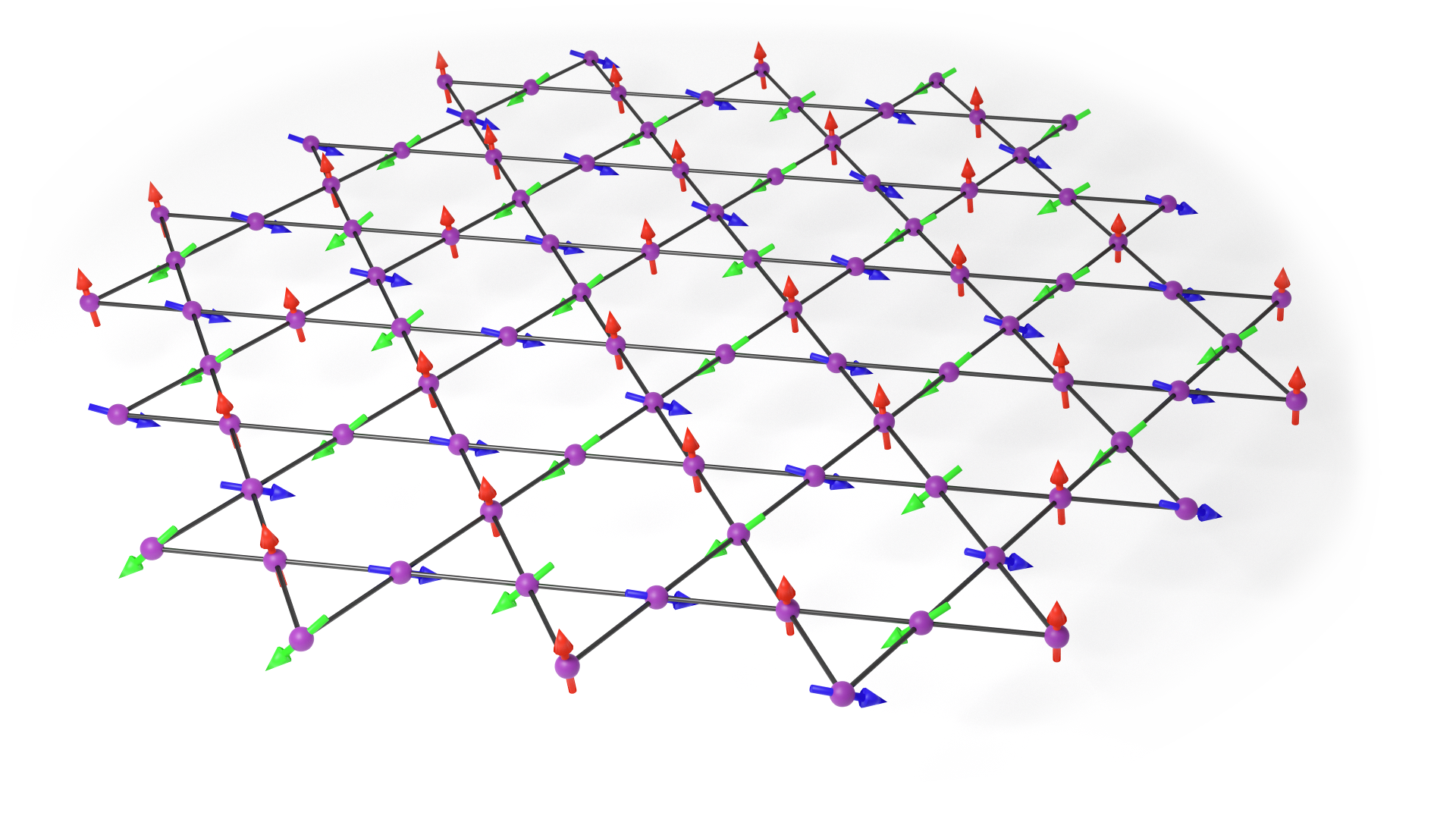}
  \caption{A kagome lattice of Heisenberg spins coupled by antiferromagnetic interaction.
  At low temperature, spins in each trianglar plaquette obey the ground-state constraint 
  $\mathbf{S}^{\rm (1)} + \mathbf{S}^{\rm (2)} + \mathbf{S}^{\rm (3)} = \mathbf{0}$. By the mechanism known as order-by-disorder a hidden biaxial spin-nematic order characterized by rank-$2$ and rank-$3$ tensors can develop at even lower temperatures.}
  \label{fig:lattice}
\end{figure}

The kagome Heisenberg antiferromagnet (KHAFM) is one of the most characteristic many-body systems where a simple Hamiltonian hosts strikingly rich physics.
A kagome lattice is built from corner-shared triangular plaquettes, as illustrated in Figure~\ref{fig:lattice}.
By placing Heisenberg spins on each lattice site and coupling the nearest-neighboring spins, one defines a kagome Heisenberg model,
$H = J \sum_{\mean{ij}} \mathbf{S}_i \mathbf{S}_j$.
When the interaction is antiferromagnetic, $J > 0$, the system is highly geometrically frustrated and fails to find a unique ground state~\cite{BookLacroix}. 
Such frustration has played a crucial role in the search for spin liquids and other exotic states of matter~\cite{Balents10}.

In the most general formulation of the problem one takes the interplay between geometric frustration, quantum fluctuations, and  thermal fluctuations into account.
However, it turns out that the thermodynamics of the classical KHAFM is already quite rewarding. 
In Ref.~\cite{Chalker92}, Chalker \emph{et al.} realized that there exists a hidden quadrupolar order characterized by a rank-$2$ tensorial order parameter.
The emergence of this order is driven by an order-by-disorder phenomenon~\cite{Villain79, Villain80}, where coplanar states are entropically selected at low temperatures owing to the presence of soft modes~\cite{Chalker92}.
Shortly afterwards, Reimers and Berlinsky~\cite{Reimers93} carried out a thorough investigation of excitations of the soft modes and the correlations of degenerate magnetic states.
In spite of limited computational resources the authors showed, by classical Monte Carlo simulations, convincing signatures that the quadrupolar order indeed appears to be (quasi-)long ranged.
At around the same time, Huse and Rutenberg~\cite{Huse92} and Ritchey \emph{et al.}~\cite{Ritchey93} proposed that the physics in the selected plane might be effectively described by three-state Potts-like degrees of freedom, and the latter paper also discussed the associated topological defects which can support a generalized Berezinskii-Kosterlitz-Thouless (BKT) transition~\cite{Berezinskii71, Berezinskii72, Kosterlitz73}. 
This is highly non-trivial, because the BKT transition is not realized in a typical two dimensional Heisenberg magnet as the fundamental group $\pi_1\left(SO(3)/SO(2)\right) = 0$ is trivial~\cite{Mermin79, Michel80}. 
It, however, becomes possible with the effective Potts-like degrees of freedom, where the fundamental group is $\pi_1 \left(O(3)/D_{3h}\right) = \bar{D}_3$ which gives rise to gapped topological defects~\cite{Michel80}, where $D_{3h}$ ($\bar{D}_3$) denotes the (binary) dihedral group.
Although these results firmly evidenced that there is further structure in the selected coplanar states, it was only much later that Zhitomirsky pointed out that the structure is described by another hidden order, a rank-$3$ octupolar order~\cite{Zhitomirsky02, Zhitomirsky08}.
This author also showed that such order coexists with pinch points, which are usually seen in classical spin liquids such as spin ice~\cite{Henley10, Castelnovo12}.
While the quadrupolar and octupolar orders are firmly established, they do not exhaust the rich hierarchy of phases in the classical KHAFM; specifically, the possibility of dipolar order is debated in the literature.
Already in Ref.~\cite{Huse92}, the authors argued that fluctuations around the extensively degenerate KHAFM ground states interact in a non-linear way and will lead to unequally weighted Potts states.
As a consequence, there may exist a critical point for a long-range antiferromagnetic $\sqrt{3} \times \sqrt{3}$ order in the $T \rightarrow 0$ limit.
This scenario was further developed by Henley with a self-consistent effective Hamiltonian approach~\cite{Henley09}.
A recent Monte Carlo simulation, equipped with an ingenious cluster update, by Chern and Moessner also suggests that the $\sqrt{3} \times \sqrt{3}$ magnetization retains a finite, though remarkably small, value in the thermodynamical limit as $T \rightarrow 0$~\cite{Chern13}.

Whereas establishing the $T \rightarrow 0$ magnetic order is very difficult because of the lack of a controlled theory~\cite{Henley09} and efficient algorithms to simulate large system sizes from first principles at extremely low temperature~\cite{Chern13, Schnabel12},
a major challenge in understanding the hidden multipolar orders in the classical KHAFM is rooted in the complexity of those high-rank tensorial order parameters.
It is therefore interesting to ask whether machine learning (ML), which is designed to analyze complex patterns, can assist in the analysis of such problems.
This also concerns ML techniques as practical tools for physics. Although these techniques have proven useful for classifying phases and detecting phase transitions~\cite{Ponte17, Wang16, Nieuwenburg17, Carrasquilla17, LiuYH18, Rodriguez-Nieva19, Zhang20}, representing quantum wave functions~\cite{CarleoTroyer17,Huang17, Cai18, Melko19, Pfau19, Hermann19}, designing algorithms~\cite{Liao19, CarleoChoo19, Nagai17, Xu17}, and predicting properties of materials~\cite{Xie18, Lee16, Isayev17, Zhu18} (see Refs.~\cite{Carleo19, Schmidt19, Carrasquilla20} for recent reviews), the number of applications to intricate issues remains limited.

In the present paper, we apply our recently developed tensorial kernel support vector machine (TK-SVM)~\cite{Greitemann19, Liu19, Greitemann19b}, which is an unsupervised and interpretable approach, to the classical KHAFM.
We show that our machine readily picks up the hidden quadrupolar and octupolar orders and gives their tensorial order parameters in analytical form while its only inputs are real-space spin configurations.
It identifies the ground-state constraint (GSC) and different temperature scales in the KHAFM.
In addition, the machine also recognizes that the magnetic correlation of classical KHAFM at low-temperature is dominated by a $\sqrt{3} \times \sqrt{3}$ structure.

The manuscript is organized as follows.
In Section~\ref{sec:method}, we review the method of TK-SVM.
In Section~\ref{sec:pd}, the finite-temperature phase diagram of KHAFM is discussed.
Section~\ref{sec:op} is devoted to the multipolar order parameters and GSC.
Section~\ref{sec:sqrt3} discusses magnetic correlations.
We conclude in Section~\ref{sec:conclusion}.

\begin{figure*}
  \centering
  \begin{tabular}{c}
  \subfloat[single-spin cluster, \mbox{rank-2}]{\includegraphics{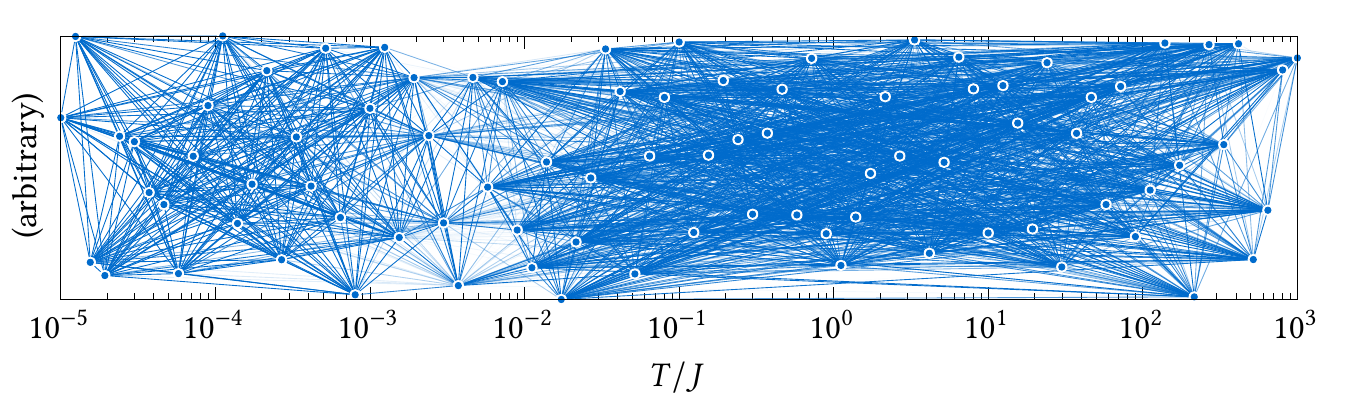}} \\
  \subfloat[single-spin cluster, \mbox{rank-3}]{\includegraphics{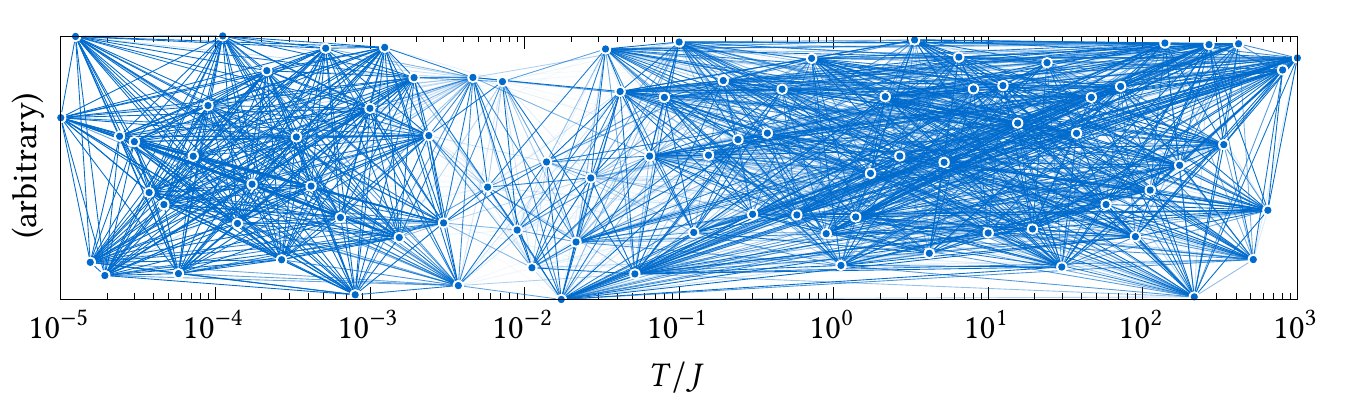}} \\
  \subfloat[three-spin triangular cluster, \mbox{rank-2}]{\includegraphics{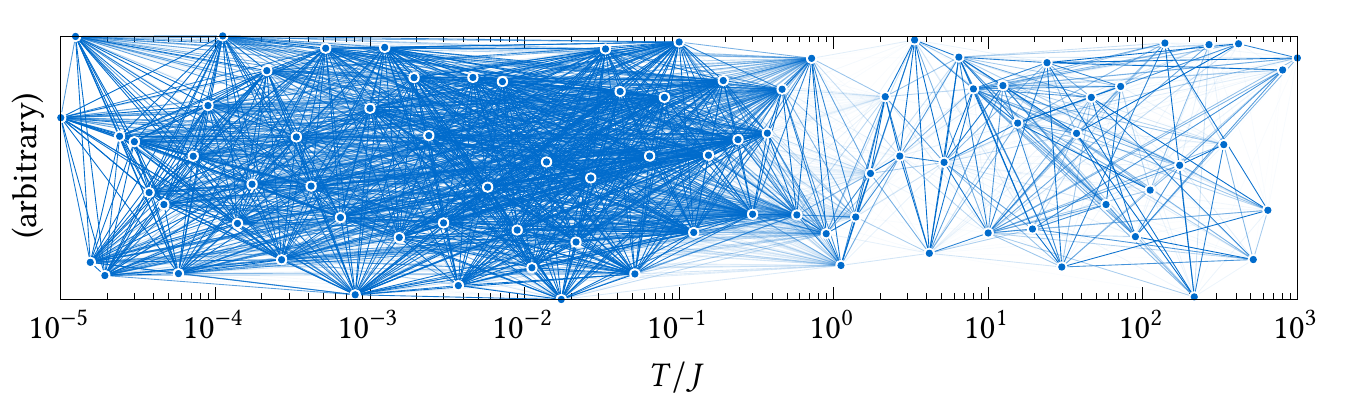}} \\
  \subfloat[three-spin triangular cluster, \mbox{rank-3}]{\includegraphics{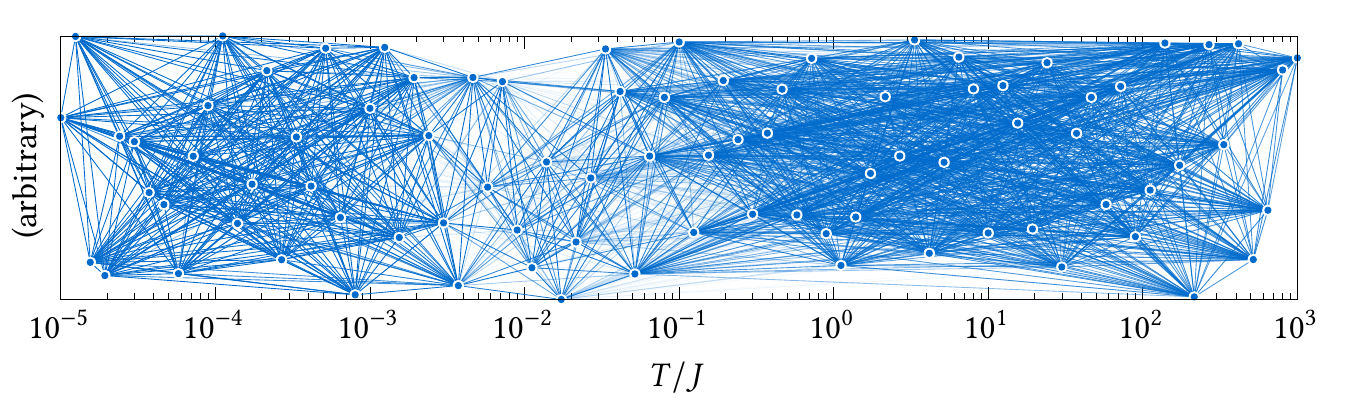}}
  \end{tabular}
  \caption{Graphs corresponding to the multiclassification along four different TK-SVM setups (two cluster choices (a kagome unit cell and a single spin) combined with the two ranks,  $n=2$ and $n=3$). Each vertex represents a different temperature $T/J$. Edges connecting the vertices are weighted by the bias parameter $\rho$ in the decision function Eq.~\eqref{eq:d(x)}; see Section~\ref{sec:gp} for details. In order to facilitate the graphical representation of the one-dimensional parameter space, a random second coordinate is introduced, which offsets the temperatures vertically and which hence avoids overlaps of edges. The clustering of the vertices is described by the respective Fiedler vector in Figure~\ref{fig:kagome_fiedler}.}
  \label{fig:kagome_graph}
\end{figure*}

\section{Tensorial-kernel support vector machine} \label{sec:method}

The tensorial kernel support vector machine (TK-SVM) is a numerical method to detect general symmetry-breaking orders~\cite{Greitemann19, Liu19} and emergent local constraints~\cite{Greitemann19b, Liu20} in the problem of phase classifications; see also Ref.~\cite{Jonas_thesis}. It inherits from support vector machines~\cite{CortesVapnik95, BookVapnik}, a well-known and successful classifying technique in machine learning, the property that it is {\it interpretable}: The decision function, which is the optimal classifier between two sets of data with a distinct property, can be shown to learn the square of order parameters when a quadratic kernel is used~\cite{Ponte17, Greitemann19, Liu19}. In TK-SVM the order parameter (or local constraint) can be any local tensor of a possibly high rank~\cite{Greitemann19, Liu19}. 

Furthermore, the decision function also contains an offset known as the bias. Specifically for phase classification, the bias is sensitive to the presence of phase transitions or crossovers~\cite{Liu19, Greitemann19b}. It can be analyzed prior to and independent of the determination of the order parameters and allows one to perform an {\it unsupervised} graph partitioning of the phase diagram. 

The main advantages of our approach are that the user does not need to devise suitable order parameters, which are typically very hard to construct for exotic states of matter, and that one gets near certainty about all phases with learned order parameters without supervision, resulting in enormous speedups for the analysis of an (unknown) phase diagram. Below we explain the most important concepts of TK-SVM.

\subsection{TK-SVM decision function} \label{sec:d(x)}

The TK-SVM finds the optimal decision function
\begin{equation}\label{eq:d(x)}
    d(\mathbf{x}) = \sum_{\mu\nu} C_{\mu\nu} \phi_\mu(\mathbf{x}) \phi_\nu(\mathbf{x}) - \rho,
\end{equation}
in classifying two sets of data with a distinct property (e.g. a different order parameter or local constraint) in the sense of determining a  maximal margin between the sets.

Here, $\mathbf{x}$ denotes a real-space configuration of $N$ spins, which serves as the training data,
\begin{equation}
    \mathbf{x} = \{S_{i,a} | a = x,y,z; i = 1,2, \dots, N\}.
\end{equation}
$O(3)$ spins are of interest to us,  but applications to $XY$ and Ising spins are straightforward.
It plays no role what algorithm has generated the samples $\mathbf{x}$.

The power of SVMs lies in the usage of appropriate kernel methods. Samples are mapped by some  transformation $\phi$  onto a feature space where only the inner product of the transformed data needs to be known~\cite{BookVapnik}. 
The conditions on $\phi$ are very mild. Often, a linear separation in feature space is possible, which can correspond to a highly non-linear separation in physical space.
In the TK-SVM, $\phi_\mu(\mathbf{x})$ maps $\mathbf{x}$ to degree-$n$ monomial configurations,
\begin{equation} \label{eq:phi}
  \bds{\phi}(\mb{x}) =\{\phi_{\mu}\}= \{ \langle S^{\alpha_1}_{a_1} \dots S^{\alpha_n}_{a_n} \rangle_{\rm cl} \},
\end{equation}
where $a_1, \dots, a_n = x,y,z$; 
$\alpha_1, \dots, \alpha_n = 1, 2, \dots,r$;
$\mu = (\alpha_1, a_1; \dots; \alpha_n, a_n)$ is a collective index;
$\langle \dots \rangle_{\rm cl}$ denotes a lattice average up to clusters of $r$ spins.
This mapping is established by the fact that local orientational orders can be generally represented by finite-rank tensors built from a finite number of vector fields~\cite{Liu16, Nissinen16, Michel01}.  
For example, magnetic orders are defined as rank-$1$ tensors, and quadrupolar orders correspond to rank-$2$ tensors.
Emergent local constraints, such as ground-state constraints for spin liquids, show up as relations between local tensors~\cite{Greitemann19b, Liu20}. 
The dimension of the $\phi$-space is $(3r)^n$. However, as it contains a massive amount of redundant information, the actual complexity in the SVM optimization problem is linearly determined by the number of independent components, given by $\frac{(3r+n-1)!}{n!(3r-1)!}$~\cite{Greitemann19}.
While this can still be a big number, a bottleneck is only encountered in extreme situations when dealing with very large clusters at high ranks.
In real applications, a TK-SVM can handle, for example, a cluster of several hundreds spins at rank-$2$ without a problem, which is only necessary when a quadrupolar order or a spin-liquid constraint has such a vast ``unit cell". In the case of rank $n=1$, which detects magnetic orders, the size of a feasible cluster can be even greater.

Coefficient matrices, $C_{\mu\nu}$, will be constructed by support vectors,
\begin{equation}\label{eq:Cmunu}
  C_{\mu\nu} = \sum_{k} \lambda_k \phi_{\mu}(\mb{x}^{(k)})\phi_{\nu} (\mb{x}^{(k)}),
\end{equation}
where $\lambda_k$ is the Lagrangian multiplier related to the $k$-th sample $x^{(k)}$ and is solved in the underlying SVM optimization problem.
The coefficient matrices can be viewed as encoders of order parameters, from which analytical expressions of the detected orders and constraints are extracted.
As we investigated in Ref.~\cite{Greitemann19} by comparing results using $N_s = 10^2$ to $10^5$ samples, a few hundred samples can already give coefficient matrices of decent quality; more samples reduce statistical errors.
The complexity of the underlying SVM optimization problem generally scales as $\mathcal{O}(N_s^2)$ to $\mathcal{O}(N_s^3)$~\cite{Bottou07}.
However, the computational cost for carrying out the machine learning is in general small compared to that of generating the training data.

The parameter $\rho$ is called the bias. In a binary classification over two sample sets $A$ and $B$, its behavior can be summarized as follows~\cite{Greitemann19b} 
\begin{align}\label{eq:rho_rules}
    \rho(A\,|\,B) \begin{cases}
        \!\begin{rcases}
          \gg 1\\
          \ll -1
        \end{rcases}\quad & \textup{$A$, $B$ in the same phase}, \\
        \approx 1 & \textup{$A$ in the disordered phase}, \\
        \approx -1 & \textup{$B$ in the disordered phase}, \\
        \in (-1,1) & \textup{not directly comparable}.
    \end{cases}
\end{align}
Therefore, the $\rho$ parameter can act as an indicator of phase transitions and crossovers.
Its magnitude indicates the absence or presence of a phase transition or crossover, 
\begin{align}\label{eq:rho_reduced}
    \abs{\rho_{\rm AB}} \begin{cases}
        \gg 1 & \textup{${\rm A, B}$ in the same phase}, \\
        \lesssim 1 & \textup{${\rm A, B}$ in different phases}.
    \end{cases}
\end{align}
The sign of $\rho$ further reveals which sample set originates from the (dis-)ordered side, namely, the orientation of the transition or crossover.
The last case in Eq.~\eqref{eq:rho_rules} corresponds to situations where $A$ and $B$ have characteristics that can not straightforwardly be compared; hence, a relative disorderedness may not be well defined.

\subsection{Graph partitioning}\label{sec:gp}

The use of graph partitioning in TK-SVM maximally exploits the reduced criterion  Eq.~\eqref{eq:rho_reduced} for $\rho$.
Consider a spin model involving a set of physical parameters such as interactions and temperature, whose  phase diagram we seek to learn.
We collect spin configurations from distinct parameter points in the physical parameter space. The distribution of the points can be uniform or not in case one wants to have a higher density of points in the regions of special interests.
We then perform SVM multi-classification over the collected data. For $M$ sets of samples, one multi-classification produces $M(M-1)/2$ decision functions; each is responsible for a binary classification between two sample sets~\cite{HsuLin02}.

The graph is constructed by viewing those parameter points as vertices and assigning an edge to each pair of them, while the edge weights are determined by the value of $\rho$ in the corresponding decision function, leading to a graph of $M$ vertices and $M(M-1)/2$ edges.

The graph may be partitioned by different means. A simple yet efficient approach is Fiedler's theory of spectral clustering~\cite{Fiedler73, Fiedler75}.
According to  Fiedler's theory, the graph can be described by a $M \times M$ Laplacian matrix $\hat{L}$.
The off-diagonal elements of $\hat{L}$ record the edge weights, and the diagonal elements represent the total weights of every vertex.
The second smallest eigenvalue, $\lambda_2$, of $\hat{L}$ measures the algebraic connectivity of the graph.
The eigenvector, $\mathbf{f}_2$, associated with $\lambda_2$ is known as the Fiedler vector and reflects the clustering of the graph.
As we shall discuss in Section~\ref{sec:pd}, the Fiedler vector plays the role of a phase diagram.

\begin{figure}
  \centering
  \includegraphics[width=0.48\textwidth]{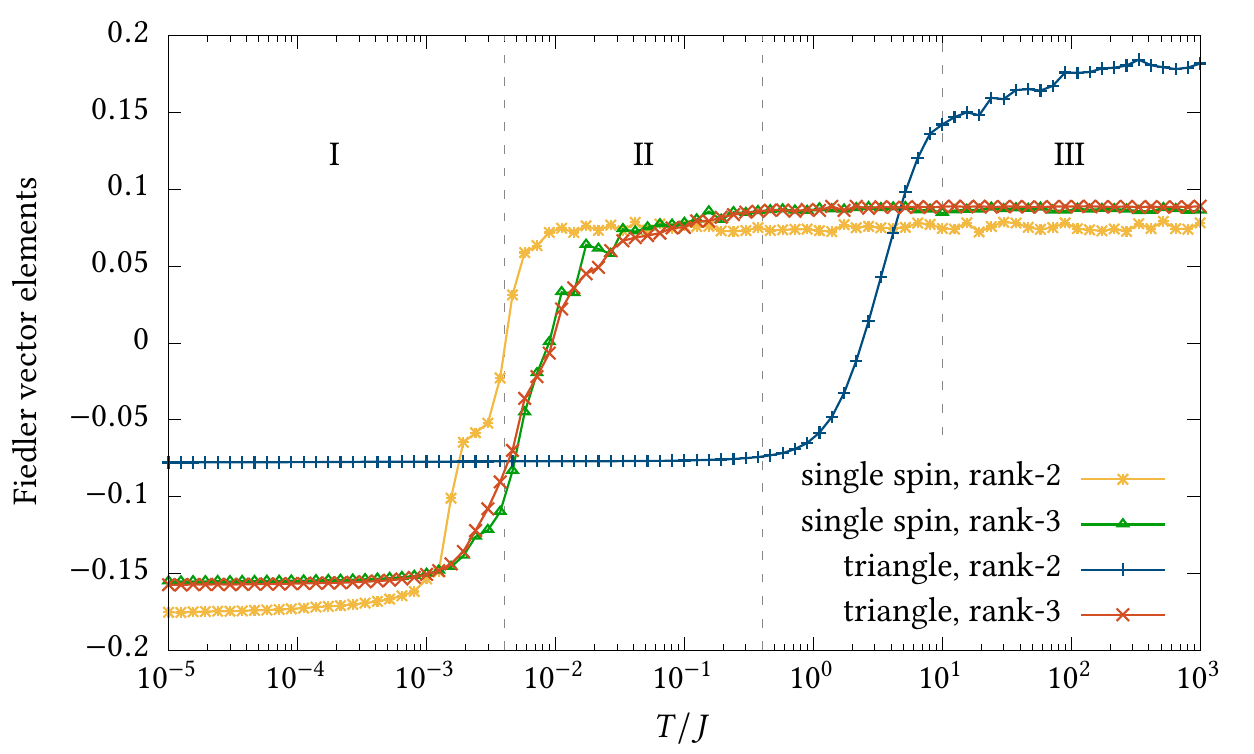}
  \caption{Fiedler vectors obtained from the partition of the four graphs. 
  The entries of each Fiedler vector are plotted against temperature.
  The curve corresponding to the triangle cluster at rank-$2$ captures the ground-state constraint Eq.~\eqref{eq:constraint}.
  The two rank-$3$ curves realize the hidden octupolar order, formulated in Eqs.~\eqref{eq:D3h_op} and ~\eqref{eq:op_oct_c3}. 
  The single-spin cluster at rank-$2$ realizes the hidden quadrupolar Eq.~\eqref{eq:Q_ab}, which occurs simultaneously with the octupolar one at $T_K = 0.004J$.
  The dashed lines indicate the boundaries of the three regimes:
  I, hidden $D_{3h}$ nematic; II, cooperative paramagnet; III, trivial paramagnet.
  The interval between $T/J = 0.4$ and $10$ is a crossover region where the ground-state constraint develops.}
  \label{fig:kagome_fiedler}
\end{figure}

\section{The phase diagram learned by TK-SVM} \label{sec:pd}

In order to learn the phase diagram of the classical KHAFM in an unsupervised way, we start the analysis with the graph partitioning.
For this purpose, we collect spin configurations ranging from high to extremely low temperatures: we choose $85$ logarithmically equidistant temperatures between $T/J = 10^3$ and $10^{-5}$, and store $1000$ independent configurations at each temperature.
The samples are obtained by classical Monte Carlo simulations utilizing parallel tempering and heat-bath updates for a lattice consisting of $3072$ spins ($32 \times 32$ kagome unit cells)~\cite{Jonas_thesis}.

Next, we apply TK-SVM between any two temperature points and repeat this for different ranks and clusters, which both need to be sufficiently large in order to accommodate the correct orders. In practice, we typically examine clusters comprising a number of lattice unit cells and scrutinize the ranks that are compatible with a crystallographic system ($n = 1,2,3,4,6$).
For the kagome Heisenberg model, it turns out that the optimal choice for learning the quadrupolar and octupolar order is the kagome unit cell (i.e., a three-spin triangle) with ranks $n=2$ and $3$.
The results will be compared with those using a single-spin cluster.
In Section~\ref{sec:sqrt3}, we will also discuss the rank $n = 1$ results of magnetic correlations.

Here are hence four different TK-SVM classification setups (the two types of clusters (a single unit cell and a single spin) combined with the  two ranks ($n=2$ and $n=3$).
Each leads to a graph of $85$ vertices and $3570$ edges, visualized by Figure~\ref{fig:kagome_graph}.
The edge weights are defined by a Lorentzian function
\begin{align}\label{eq:weight}
  w(\rho) &= 1 - \frac{\rho_c^2}{(|\rho|-1)^2+\rho_c^2} \in [0,1),
\end{align}
where $\rho$ is the bias parameter of the binary classification, and 
$\rho_c$ determines a characteristic value for ``$\gg 1$'' in the reduced $\rho$ criterion Eq.~\eqref{eq:rho_reduced}.
The choice of $\rho_c$ is nevertheless not crucial because vertices in the same phase typically have stronger connections than those from different phases.
The robustness of the graph partitioning follows from the fact that $\rho_c$  can be varied over several orders of magnitude without significant changes to the topology of the phase diagram~\cite{Greitemann19b}.

Partitioning these graphs leads to four different Fiedler vectors, depicted in Figure~\ref{fig:kagome_fiedler}.
Clearly, the temperature axis is divided into three regions:
A rank-$2$ quantity constructed from the triangle cluster (see also panel (c) in Figure~\ref{fig:kagome_graph}) discriminates around a temperature $T/J \approx 1$ a high-temperature region (region III) from the rest.
Both the single-spin and the triangle cluster (cf. panels (b) and (d) in Figure~\ref{fig:kagome_graph}) define a rank-$3$ quantity which distinguishes the very low temperatures (region I,  $T/J \le 0.004$) from the intermediate temperature region II. 
The result of the single-spin cluster at rank-$2$ (cf. panel (a) in Figure~\ref{fig:kagome_graph}) is in line with the results  the rank-$3$ ones.
The discrimination of these regions reproduces in fact the finite-temperature phase diagram of the kagome Heisenberg model~\cite{Zhitomirsky08,Chalker92} -- even before we visit the nature of those phases.

\section{Analytical order parameters} \label{sec:op}

Having learned the topology of the  KHAFM phase diagram, we now interpret the nature of each phase by their order parameters and possible local constraints.
This will be done by extracting the analytical order parameters from the $C_{\mu\nu}$ matrices.
In order to reduce the statistical errors on $C_{\mu\nu}$, we pool the data according to the phase diagram of Figure~\ref{fig:kagome_fiedler}.
In addition, we introduce $25,\!000$ fictitious configurations generated at $T = + \infty$ which represent completely disordered states.
This sets up a reduced multi-classification problem among four classes: regime I ($28,\!000$ samples), regime II ($21,\!000$ samples), regime III ($21,\!000$ samples), and regime $T_{\infty}$.
Consequently, we obtain six $C_{\mu\nu}(A\,|\,B)$ for each rank and cluster, with $A, B \in \{ {\rm I, II, III, T_{\infty} }\}$~\cite{Jonas_thesis}.

Not surprisingly, the high-temperature regime III is not distinct from the $T_{\infty}$ regime; the associated coefficient matrices are noise-like.
Hence, only the two low-temperature regimes need further interpretation.

\begin{table*}[ht!]
  \centering
  \small
  \begin{tabular*}{0.9\textwidth}{@{\extracolsep{\fill}}lcccccccc}
    \toprule
    & & \multicolumn{1}{c}{\kern-0.5em$p[Q_{x^2+y^2+z^2}]$} & \multicolumn{1}{c}{$p[Q_{z^2}]$} & \multicolumn{1}{c}{\kern-0.2em$p[Q_{x^2+y^2}]$} & \multicolumn{1}{c}{$p[Q_{x^2-y^2}]$} & \multicolumn{1}{c}{$p[Q_{xy+yx}]$} & \multicolumn{1}{c}{\kern-0.2em$p[Q_{yz+zy}]$} & \multicolumn{1}{c}{\kern-0.2em$p[Q_{zx+xz}]$}\\\midrule
                    & on-site & -0.487 &  1.460 &  0.731 &  0.732 &  0.733 &  0.727 &  0.735\\
    I $|$ II          & cross   &  0.081 & -0.730 & -0.366 & -0.366 & -0.367 & -0.363 & -0.367\\
                    & bond    &  0.038 &  0.365 &  0.183 &  0.183 &  0.183 &  0.182 &  0.184\\\midrule
                    & on-site &  0.059 & -0.176 & -0.088 & -0.088 & -0.088 & -0.088 & -0.088\\
    I $|$ $T_\infty$  & cross   &  0.910 &  0.089 &  0.044 &  0.044 &  0.044 &  0.044 &  0.044\\
                    & bond    & -0.462 & -0.044 & -0.022 & -0.022 & -0.022 & -0.022 & -0.022\\\midrule
                    & on-site &  0.000 &  0.000 &  0.000 &  0.000 &  0.000 &  0.000 &  0.000\\
    II $|$ $T_\infty$ & cross   & -0.998 & -0.001 &  0.000 &  0.000 &  0.000 &  0.000 &  0.000\\
                    & bond    &  0.445 & -0.001 &  0.000 &  0.000 &  0.000 &  0.000 &  0.000\\
    \bottomrule
  \end{tabular*}
 \caption{The weights of quadrupolar ordering components, $p[Q_\bullet]$, extracted from the corresponding coefficient matrices.
   	These weights are calculated through a least-squares fit based on all blocks in the full coefficient matrix of each of the site-site (``on-site''), site-bond (``cross''), and bond-bond (``bond'') types.
  The ratios between these weights for the three block types are given in Table~\ref{tab:gamma}.}
  \label{tab:components}
\end{table*}

\begin{table*}[ht!]
  \centering
  \begin{tabular*}{0.9\textwidth}{@{\extracolsep{\fill}}ccccccccc}
  \toprule
  & & \multicolumn{1}{c}{$\gamma_{x^2+y^2+z^2}$} & \multicolumn{1}{c}{$\gamma_{z^2}$} & \multicolumn{1}{c}{$\gamma_{x^2+y^2}$} & \multicolumn{1}{c}{$\gamma_{x^2-y^2}$} & \multicolumn{1}{c}{$\gamma_{xy+yx}$} & \multicolumn{1}{c}{$\gamma_{yz+zy}$} & \multicolumn{1}{c}{$\gamma_{zx+xz}$}\\\midrule
  \multirow{2}{*}{I $|$ II}         & c $\div$ s & -0.166 & -0.500 & -0.500 & -0.500 & -0.500 & -0.500 & -0.500\\
                                  & b $\div$ c &  0.471 & -0.500 & -0.500 & -0.500 & -0.500 & -0.500 & -0.500\\\midrule
  \multirow{2}{*}{I $|$ $T_\infty$} & c $\div$ s &        & -0.504 & -0.499 & -0.502 & -0.502 & -0.502 & -0.502\\
                                  & b $\div$ c & -0.508 & -0.492 & -0.495 & -0.493 & -0.493 & -0.493 & -0.493\\\midrule
  II $|$ $T_\infty$                 & b $\div$ c & -0.446 &        &        &        &        &        &       \\
  \bottomrule
  \end{tabular*}
  \caption{Ratios of the quadrupolar moments in Table~\ref{tab:gamma}, $\gamma_\bullet\coloneqq \mean{Q^{\alpha\beta}_\bullet}/\mean{Q^{\alpha\alpha}_\bullet}$.
  These can be calculated by taking the ratio between the weights of the relevant component with respect to ``cross'' and ``on-site'' blocks (c $\div$ s), or ``bond'' and ``cross'' blocks (b $\div$ c).}
  \label{tab:gamma}
\end{table*}

\begin{figure}
  \centering
  \includegraphics[width=0.5\textwidth]{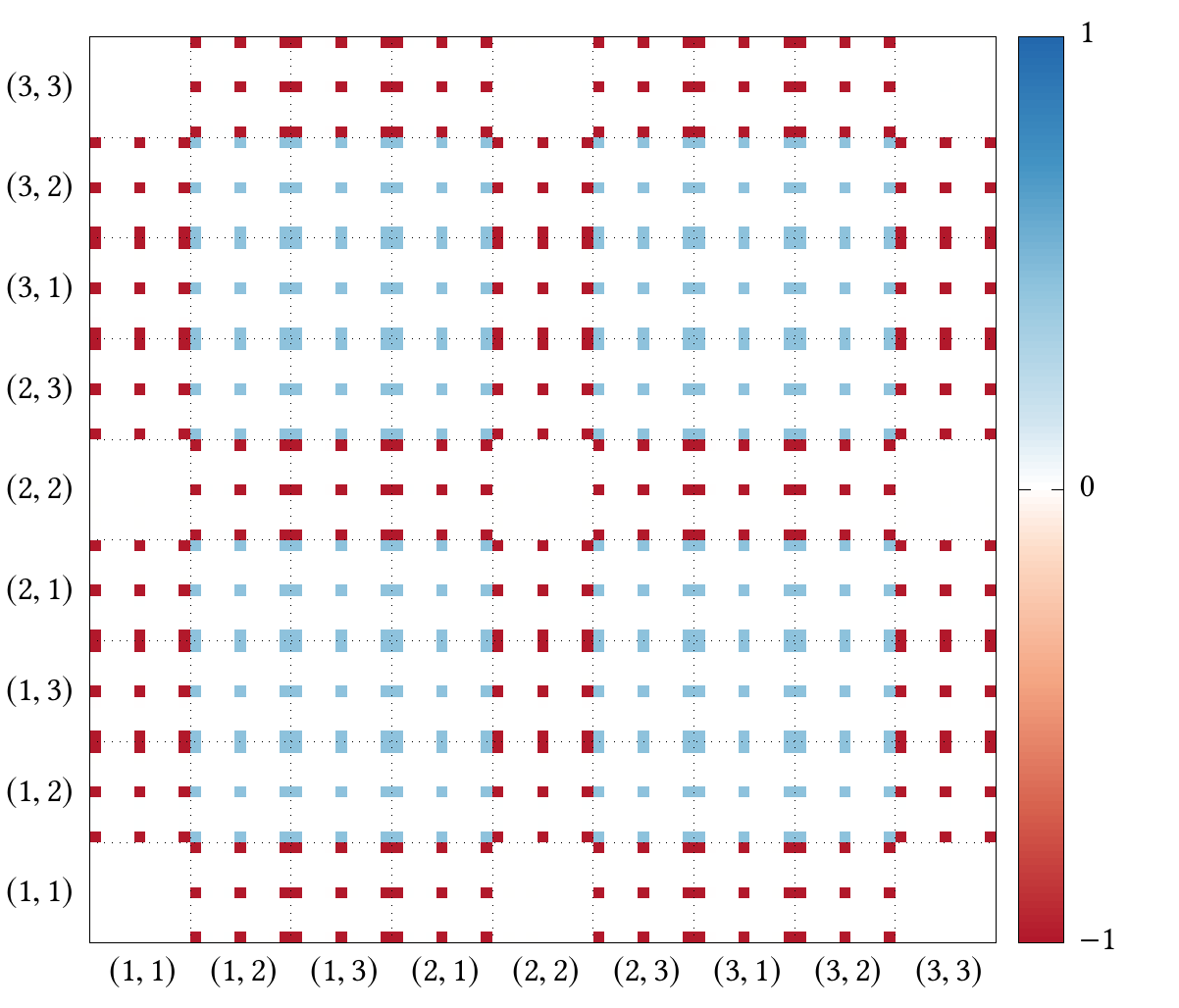}
  \caption{The coefficient matrix $C_{\mu\nu}(\textup{II}\,|\,T_\infty)$ learned from a \mbox{rank-2} TK-SVM on a three-spin triangular cluster.
  The dotted lines demarcate $9\times 9$ blocks corresponding to identical spin indices $(\alpha, \beta), (\alpha', \beta')$ while the contents of each $9\times 9$ block is enumerated in a similar way by component indices $(a,b), (a',b')$.
  On-site blocks ($\alpha=\beta, \alpha'=\beta'$) are seen to be empty, while ``cross'' and ``bond'' blocks exhibit a non-vanishing $\delta_{ab}\delta_{a'b'}$ pattern with relative weights of $-1$ and $0.446=-\gamma$, respectively. 
  See also Tables~\ref{tab:components} and~\ref{tab:gamma}.}
  \label{fig:constraint}
\end{figure}

\subsection{Ground-state constraint}\label{sec:gsc}
We first discuss the emergent local constraint in the two low-temperature phases, which drives the system to coplanar order and is manifest using the triangular cluster at rank-$2$.
The term ``emergent'' in the current case is intended to distinguish from the trivial constraint of spin normalization $|\mathbf{S}| = 1$. Examples of learning more non-trivial constraints, including previously unknown ones, can be found in Ref.~\cite{Liu20} in cases of Kitaev and $\Gamma$ spin liquids.

The coefficient matrix $C_{\mu\nu}({\rm II\,|\,T_{\infty} })$ is visualized in Figure~\ref{fig:constraint}.
It is composed of $9 \times 9$ blocks enumerated by the spin indices  $[\alpha,\beta;\alpha'\beta']$, with $\alpha, \beta = 1,2,3$.
One easily distinguishes three types of blocks: ``on-site'' ($\alpha = \beta, \, \alpha' = \beta'$), ``bond'' ($\alpha \neq \beta, \, \alpha' \neq \beta'$), and the mixed case ``cross''.
These blocks can be expressed as
\begin{align}\label{eq:C_con_block}
    C^{\alpha\beta;\alpha'\beta'}_{ab;a'b'} = \left[(\gamma+\bar\gamma\delta_{\alpha\beta})(\gamma+\bar\gamma\delta_{\alpha'\beta'})-\delta_{\alpha\beta}\delta_{\alpha'\beta'}\right]\delta_{ab}\delta_{a'b'},
\end{align}
 where $\gamma$ denotes the ratio between the weight of the on-site and bond quadratic correlations in a coefficient matrix (cf. Table~\ref{tab:components} and Table~\ref{tab:gamma}), and $\bar\gamma\coloneqq 1-\gamma$.

Substituting $C^{\alpha\beta;\alpha'\beta'}_{ab;a'b'}$ into the decision function Eq.~\eqref{eq:d(x)}, one obtains
\begin{align} \label{eq:decfun_constraint}
d(\{\vekk S_i\}) &\sim \sum_{\substack{\alpha, \beta\\\alpha',\beta'}}\sum_{\substack{a, b\\a', b'}}C^{\alpha\beta;\alpha'\beta'}_{ab;a'b'}\mean{S^\alpha_aS^\beta_b}\mean{S^{\alpha'}_{a'}S^{\beta'}_{b'}} \nonumber \\
  &= \bigg(\gamma \mean{\sum_a\big(\sum_\alpha S^\alpha_a\big)^2} + \bar\gamma \sum_\alpha \mean{\left\|\vekk S^\alpha\right\|^2 } \bigg)^2 \nonumber \\
      & \qquad - \bigg(\sum_\alpha\mean{\left\|\vekk S^\alpha\right\|^2}\bigg)^2 \nonumber  \\
  &= 9\bigg[\gamma^2\big(\Gamma-\frac{1}{\gamma}\big)^2 - 1\bigg],
\end{align}
where $\Gamma$ is a normalized constraint order parameter whose meaning will become transparent in the forthcoming discussion,
\begin{align}\label{eq:Gamma}
    \Gamma = 1 - \frac{1}{3}\mean{\big\|\sum_\alpha\vekk S^\alpha\big\|^2}_\textup{cl}. 
\end{align}

As seen in Figure~\ref{fig:constraint} and Table~\ref{tab:components}, the only non-vanishing quadratic correlation in $C_{\mu\nu}(\textup{II}\,|\,T_\infty)$ is
\begin{align}
    Q^{\alpha\beta}_{x^2+y^2+z^2} = S^\alpha_x S^\beta_x + S^\alpha_y S^\beta_y + S^\alpha_z S^\beta_z.
\end{align}
$\gamma$ is then determined by 
$Q^{\alpha \neq \beta}_{x^2+y^2+z^2} / Q^{\alpha=\beta}_{x^2+y^2+z^2} \sim -0.446$ for regime II.

The value of $\gamma$ in fact appears to be temperature-dependent and converges to $-\frac{1}{2}$ in regime I.
This can be understood from the constraint order parameter, Eq.~\eqref{eq:Gamma}.
As the squared sum in $\Gamma$ comprises three on-site ($\alpha=\beta$) and six bond ($\alpha \neq \beta$) correlations, and that their ratio is $\gamma$,
the fulfillment of $\gamma = -\frac{1}{2}$ is equivalent to the relation
\begin{align}
    \mean{\big\| \vekk{S}^{(1)} + \vekk{S}^{(2)} + \vekk{S}^{(3)} \big\|^2}_\textup{cl} = \vekk{0}.
\end{align}
Since $\big\| \dots \big\|^2$ is semi-positive definite, this in turn means a local constraint at each triangular plaquette,
\begin{align}\label{eq:constraint}
    \vekk{S}^{(1)} + \vekk{S}^{(2)} + \vekk{S}^{(3)} = \vekk{0}.
\end{align}

Namely, $\gamma = -\frac{1}{2}$, or equivalently $\Gamma = 1$, expresses the ground-state constraint of the KHAFM (the spins of every triangle lie in a plane), while the deviation of $\gamma \sim -0.446$ in the higher-temperature regime II reflects thermal fluctuations of the constraint.

 \begin{figure}[t]
 \centering
     \begin{tabular}{c}
         \subfloat[$C_{\mu\nu}({\rm I \,|\, T_{\infty}})$, single spin, rank-$2$]{\includegraphics[width=0.45\textwidth]{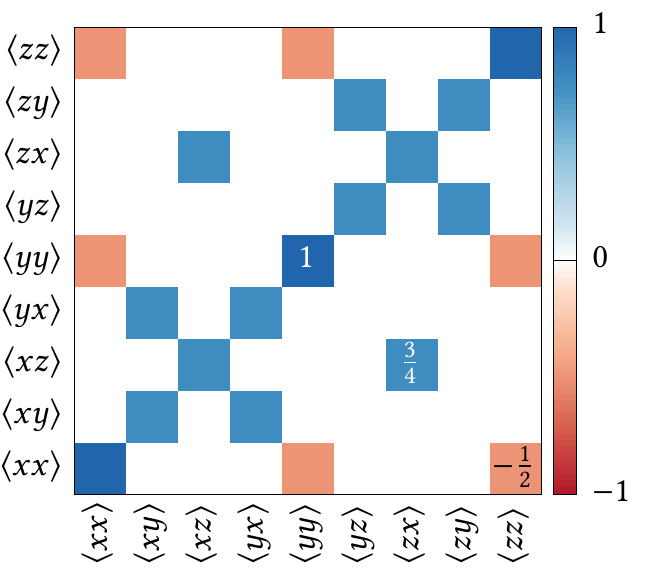}} \\
          \subfloat[$C_{\mu\nu}({\rm I \,|\, T_{\infty}})$, single spin, rank-$3$]{\includegraphics[width=0.43\textwidth]{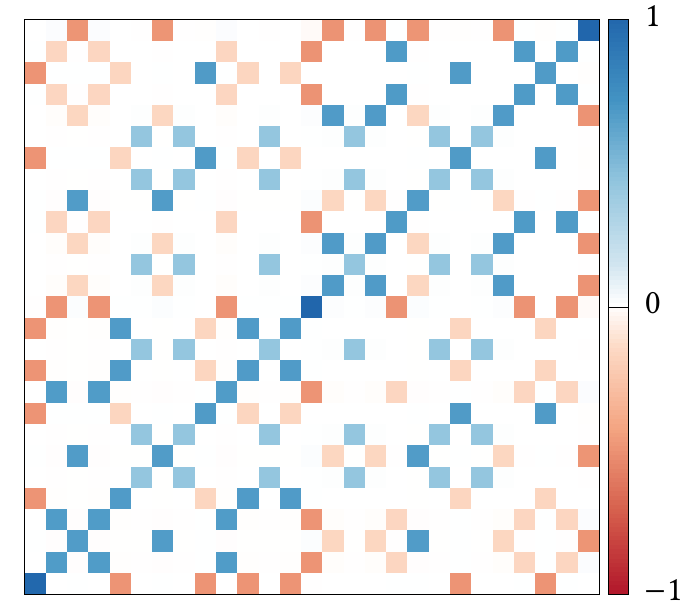}}    
     \end{tabular}
 \caption{Coefficient matrices for regime I learned from a single-spin cluster at rank-$2$ and $3$. They represent the  quadrupolar (Eq.~\eqref{eq:Q_ab}) and octupolar (Eq.~\eqref{eq:D3h_op}) tensor order parameters, respectively. They also appear to be the motifs for the blocks shown in Figs.~\ref{fig:C_c3r2} and~\ref{fig:C_c3r3} obtained with a triangular cluster.}
 \label{fig:C_c1}
 \end{figure}

\subsection{Hidden nematic order}\label{sec:nematic}
We proceed by examining the order parameter of regime I.
Given the Fiedler vectors in Figure~\ref{fig:kagome_fiedler}, it is evident that the $C_{\mu\nu}$ matrices learned with the single-spin cluster are able to distinguish regime I from the high-temperature phases.
We now look at the corresponding coefficient matrices, and will afterwards revisit this issue with the three-spin triangular cluster.

The patterns of $C_{\mu\nu}({\rm I \,|\, T_{\infty}})$ using the single-spin cluster are shown in Figure~\ref{fig:C_c1}, where it is a $9\times 9$ matrix for rank-$2$ and $27\times 27$ matrix for rank-$3$.

The rank-$2$ pattern, shown in Figure~\ref{fig:C_c1}{a},  has the following analytic expression,
\begin{align}\label{eq:C_c1r2}
  \frac{3}{4} \delta_{a a^\prime} \delta_{b b^\prime} + \frac{3}{4} \delta_{a b^\prime} \delta_{b a^\prime} - \frac{1}{2} \delta_{a b} \delta_{a^\prime b^\prime}.
\end{align}
Substituting this into the decision function Eq.~\eqref{eq:d(x)}, we obtain
\begin{align} \label{eq:Dinfh_Q}
    d(\mb{x}) &\sim \f{3}{2}\sum_{ab} \left(\left<  S_a S_b \right> - \f{1}{3} \delta_{ab}\right)^2 \nonumber \\
               &= \frac{3}{2}\Tr\left[Q_{ab} Q_{ba}\right],
\end{align}
where
\begin{align}\label{eq:Q_ab}
    Q_{ab} = \left< S_a S_b \right> - \f{1}{3} \delta_{ab}
\end{align}
is the famous uniaxial nematic tensor~\cite{BookDeGennes}. This shows that a rank-$2$ $C_{\mu\nu}({\rm I \,|\, T_{\infty}})$ using a single-spin cluster probes the hidden quadrupolar order in the KHAFM~\cite{Chalker92,Zhitomirsky08}.

The rank-$3$ pattern Figure~\ref{fig:C_c1}{b} is interpreted in the same way.
It can be expressed as 
\begin{align}\label{eq:C_c1r3}
  \delta_{aa^{\prime}} \delta_{bb^{\prime}} \delta_{cc^{\prime}} - \f{1}{5}(\delta_{aa^{\prime}} \delta_{bc} \delta_{b^{\prime}c^{\prime}} 
     + \delta_{ac} \delta_{bb^{\prime}} \delta_{a^{\prime}c^{\prime}}
    +\delta_{ab} \delta_{a^{\prime}b^{\prime}} \delta_{cc^{\prime}}),
\end{align}
leading to a rank-$3$ tensor
\begin{align} \label{eq:D3h_op}
    T_{abc} =  S_a S_b S_c - \f{1}{5} S_a \delta_{bc} - \f{1}{5} S_b \delta_{ca} - \f{1}{5} S_c \delta_{ab},
\end{align}
which is precisely the octupolar order parameter~\cite{Fel95}.

We are left with the examination of the three-spin cluster.
Although it is not quite visible from the Fiedler vector in Figure~\ref{fig:kagome_fiedler}, TK-SVM with a triangular cluster at rank-$2$ does in fact discriminate between regime I and II, but 
the distinction is blurred by the constraint, Eq.~\eqref{eq:Gamma}, which appears as the primary order parameter:
The triangular rank-$2$ curve in Figure~\ref{fig:kagome_fiedler} varies in the low-temperature regions, but the gradient is much more modest than the  profound change seen when the constraint emerges.

\begin{figure}[t]
 \centering
 	\includegraphics[width=0.5\textwidth]{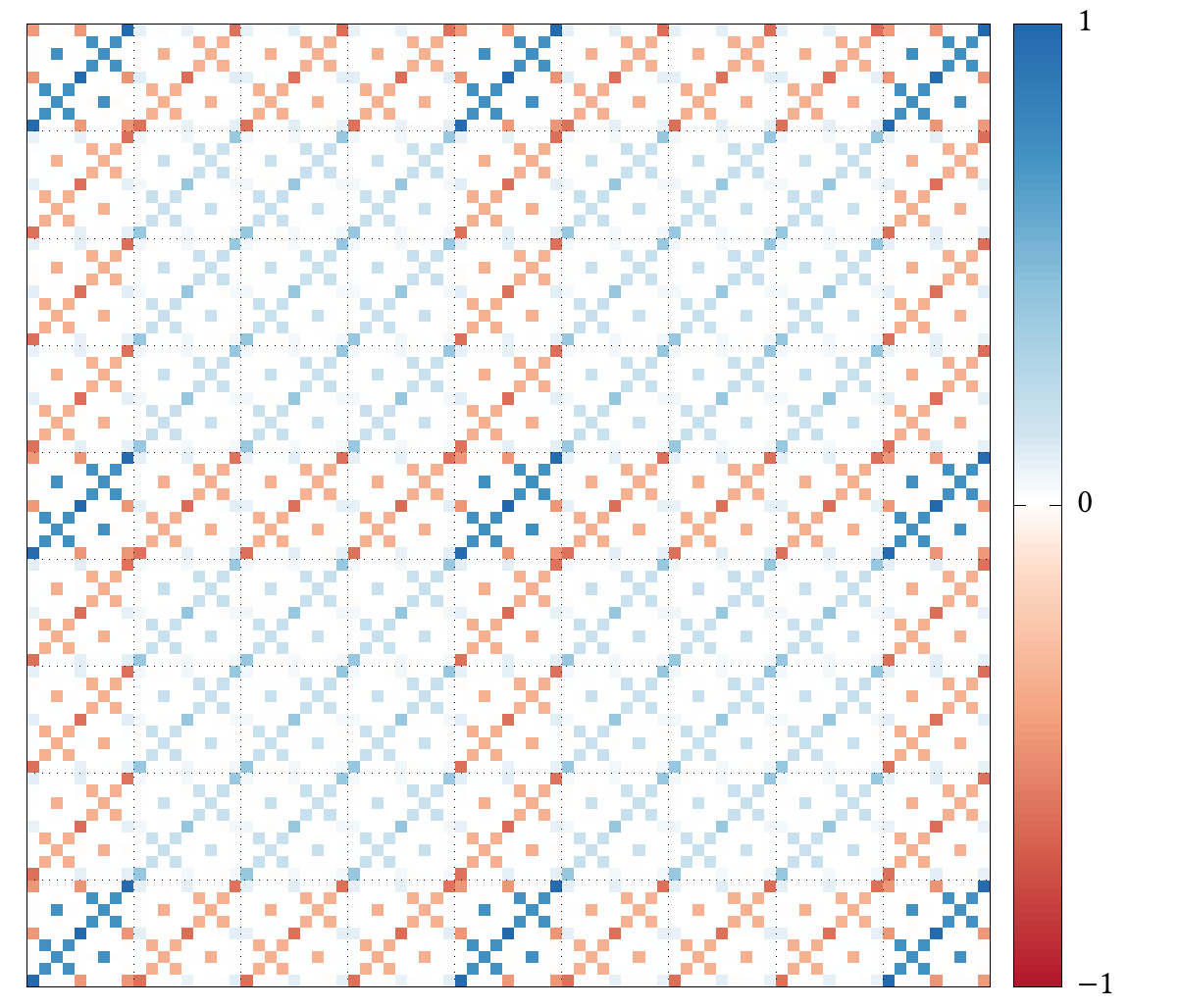}
 \caption{The coefficient matrix $C_{\mu\nu}(\textup{I}\,|\,\textup{II})$ learned from TK-SVM set up at \mbox{rank-2}  on a three-spin triangular cluster.
  The dotted lines demarcate $9\times 9$ blocks corresponding to identical spin indices.
  The same motif repeats in each block, but multiplied by a factor of $\gamma=-1/2$ on cross-type blocks and $\gamma^2$ on bond-type blocks, as compared to on-site blocks (cf. Table~\ref{tab:gamma}). This modulation encodes the constraint, in addition to the quadrupolar order parameter that is seen using a single-spin cluster.}
 \label{fig:C_c3r2}
 \end{figure}

Figure~\ref{fig:C_c3r2} shows the  $C_{\mu\nu}({\rm I \,|\, II})$ obtained from a three-spin cluster at rank-$2$, which displays a richer structure than the $C_{\mu\nu}({\rm II \,|\, T_{\infty}})$  shown in  Figure~\ref{fig:constraint}.
Following a similar analysis, it gives rise to another quadrupolar tensor,
\begin{align}\label{eq:op_quad}
  \mathbb{O}_\textup{quad} = \bigg\langle\frac{1}{3} \sum_\alpha \vekk S^\alpha\otimes\vekk S^\alpha - \big(\frac{1}{3}\sum_\alpha\vekk S^\alpha\big)^{\otimes 2}\bigg\rangle_\textup{cl}.
\end{align}
The first term is equivalent to the single-spin nematic tensor Eq.~\eqref{eq:Q_ab}, while the second term expresses the constraint, Eq.~\eqref{eq:constraint}.
In other words, the triangular cluster at rank-$2$ simultaneously detects the hidden quadrupolar order and the ground-state constraint.

\begin{figure}[t]
\centering
 	\includegraphics[width=0.5\textwidth]{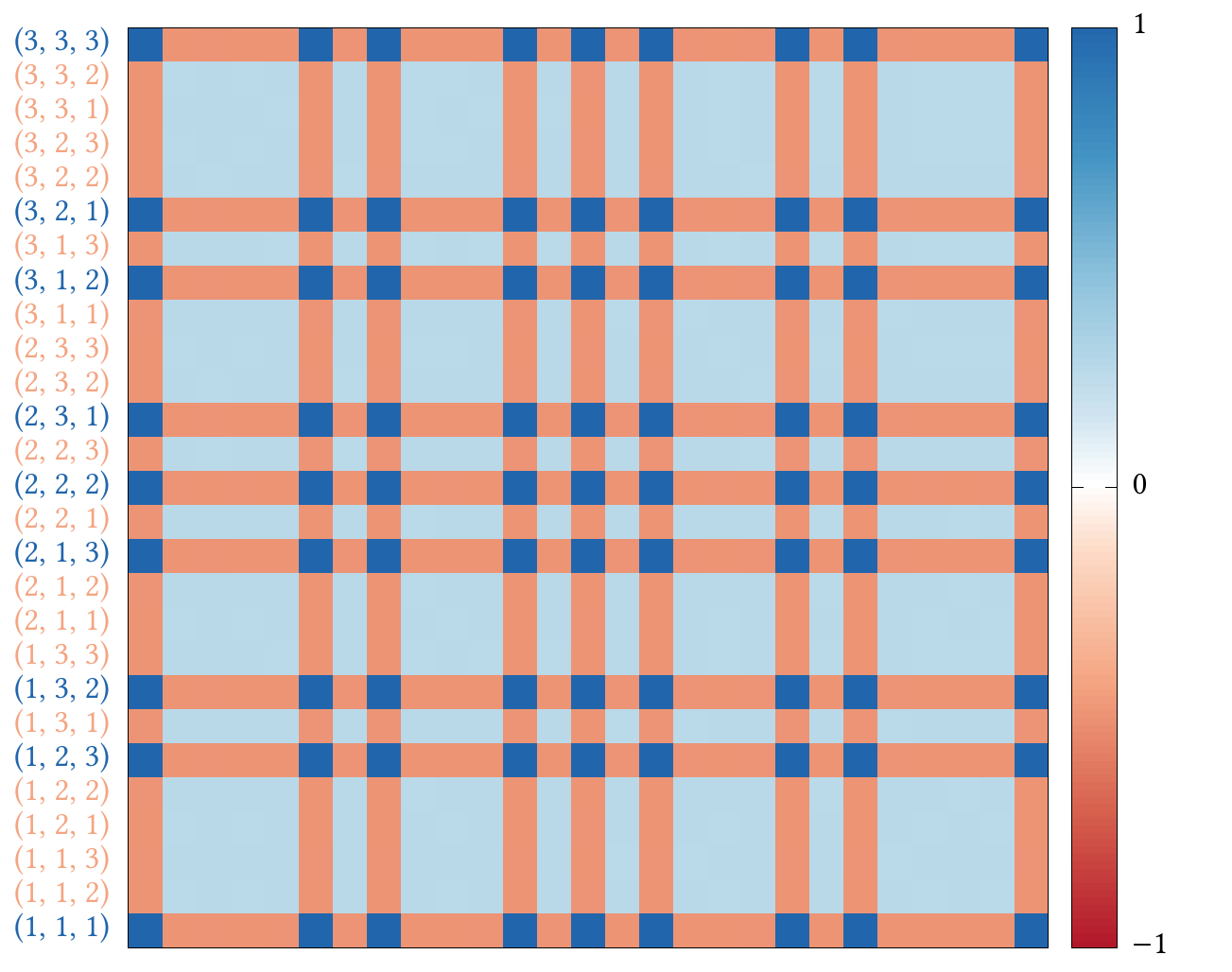}
\caption{The block structure of the coefficient matrix obtained from \mbox{rank-3} TK-SVM on a triangular cluster.
  Each block follows the form of Figure~\ref{fig:C_c1}{b}, multiplied by the block weights indicated here.
  Basis tensors $\vekk S^\alpha\otimes\vekk S^\beta\otimes\vekk S^\gamma$ with $\alpha=\beta=\gamma$ and $\alpha\neq\beta\neq\gamma$ are seen to contribute equally, whereas all others are diminished by by a factor of $\gamma=-1/2$.}
 \label{fig:C_c3r3}
 \end{figure}

In Figure~\ref{fig:C_c3r3}, we show the block structure of the rank-$3$ $C_{\mu\nu}({\rm I \,|\, T_{\infty}})$ using the triangular cluster.
The full matrix has a dimensionality of $729 \times 729$, which is composed of $27 \times 27$ blocks.
Each block exhibits the same pattern as that in Figure~\ref{fig:C_c1}{b} for the single-spin cluster, but is further multiplied with  a weight.
Its analytic expression displays a new octupolar tensor, 

\begin{align}\label{eq:op_oct_c3}
  \mathbb{O}_\textup{oct} &= \bigg\langle\frac{1}{9}\bigg[\sum_\alpha\left(\vekk S^\alpha\right)^{\otimes\,3} + \sum_{\alpha\neq\beta\neq\gamma}\vekk S^\alpha\otimes\vekk S^\beta\otimes\vekk S^\gamma\bigg] \nonumber \\
      & \qquad - \big(\frac{1}{3}\sum_\alpha \vekk S^\alpha\big)^{\otimes\,3}\bigg\rangle_\textup{cl}.
\end{align}
The first term of $\mathbb{O}_\textup{oct}$ reproduces the single-spin order parameter Eq.~\eqref{eq:D3h_op}.
The second term represents the ground-state constraint Eq.~\eqref{eq:constraint} and will become a constant tensor, $\mathbf{0}^{\otimes 3}$, when  $\Gamma$ saturates to $\Gamma = 1$, which is the case in regime I.  
The second term is a rank-$3$ tensor defined by the three spins in a triangular plaquette.
It serves as an alternative characterization of the octupolar order and is equivalent to the single-spin form.
This also explains the agreement of the two rank-$3$ Fiedler vectors in Figure~\ref{fig:kagome_fiedler}.

\begin{table*}[ht!]
  \centering
  \renewcommand{\arraystretch}{1.2}
  \begin{tabular*}{0.9\textwidth}{@{\extracolsep{\fill}}ccccc}
    \br
    \multicolumn{1}{c}{Cluster} & \multicolumn{2}{c}{single spin} & \multicolumn{2}{c}{triangle}\\
    \cmidrule(lr){2-3}\cmidrule(lr){4-5}
    \multicolumn{1}{c}{Rank} & \multicolumn{1}{c}{2} & \multicolumn{1}{c}{3} & \multicolumn{1}{c}{2} & \multicolumn{1}{c}{3}\\\midrule
    I $|$ II           & -1.0131 & -1.0131 & -3.286  & -1.0145 \\
    I $|$ III          & -1.0121 & -1.0041 & -1.0035 & -1.0026 \\
    I $|$ $T_\infty$   & -1.0129 & -1.0044 & -0.9928 & -1.0028 \\
    II $|$ III         & 10.63   & -0.9592 & -1.0260 & -1.0691 \\
    II $|$ $T_\infty$  &  4.218  & -1.0598 & -1.0103 & -1.0805 \\
    III $|$ $T_\infty$ &  2.158  & -1285   & -20.54  & -1.9802 \\
    \br
  \end{tabular*}
  \caption{TK-SVM biases obtained from the reduced multiclassification problems among the temperature regimes revealed by Figure~\ref{fig:kagome_fiedler} as well as the fictitious control group, $T_\infty$. The meaning of these values is understood by the $\rho$ criterion Eq.~\eqref{eq:rho_rules}.}
  \label{tab:bias}
\end{table*}

\subsection{Phase Hierarchy}\label{sec:hierarchy}
Let us put the learned order parameters and constraints together and infer a single coherent physical picture. Regime III is a trivial paramagnet (PM) which is equivalent to the infinite temperature state. Regime II is an instance of a cooperative paramagnet (CPM), which may also be referred to as a classical spin liquid and is characterized by an emergent local constraint. Regime I meets the standard of a biaxial spin nematic (BSN) where the uniaxial quadrupolar and the biaxial octupolar order together define the $D_{3h}$ symmetry~\cite{Nissinen16}. 

Accordingly, the temperature scales in the phase diagram Figure~\ref{fig:kagome_fiedler} can be understood from a hierarchy of disorderedness,
\begin{equation}\label{eq:hierarchy}
\underset{\text{PM (III)}}{\mathrm{O}(3)} \longrightarrow
		\underset{\text{CPM (II)}}{\mathrm{O}(3)}
				\longrightarrow
					\underset{\text{BSN (I)}}{D_{3h}}.
\end{equation}
In the high-temperature PM phase, spins can freely fluctuate. In the intermediate CPM phase, although the system still preserves the $O(3)$ symmetry of the KHAFM Hamiltonian, fluctuations of spins become correlated.
Then the BSN order emerges from a constrained subset of the phase space.

This phase hierarchy is also reflected by the bias parameters in the four different multi-classification setups, which are shown in Table~\ref{tab:bias}.
Following the bias criterion of Eq.~\eqref{eq:rho_rules}, regime I (BSN) has the least disorder since $\rho({\rm I\,|\,II, \,III}) \lesssim -1$ in all instances.
The $\rho(\textup{I}\,|\,\textup{II})$ learned with the triangular cluster at rank-$2$ acquired a value noticeably smaller than $-1$, since the constraint Eq.~\ref{eq:constraint} is satisfied in both phases ($\Gamma \sim 1$).
However, its sign remains revealing that regime II is more disordered.   
Regime III (PM) is the most disordered as $\rho({\rm I, \, II\,|\, III}) \lesssim -1$ in all cases, with one exception as the single-spin cluster cannot represent the three-spin constraint Eq.~\eqref{eq:constraint}.

\begin{figure}
  \centering
  \includegraphics[width=0.45\textwidth]{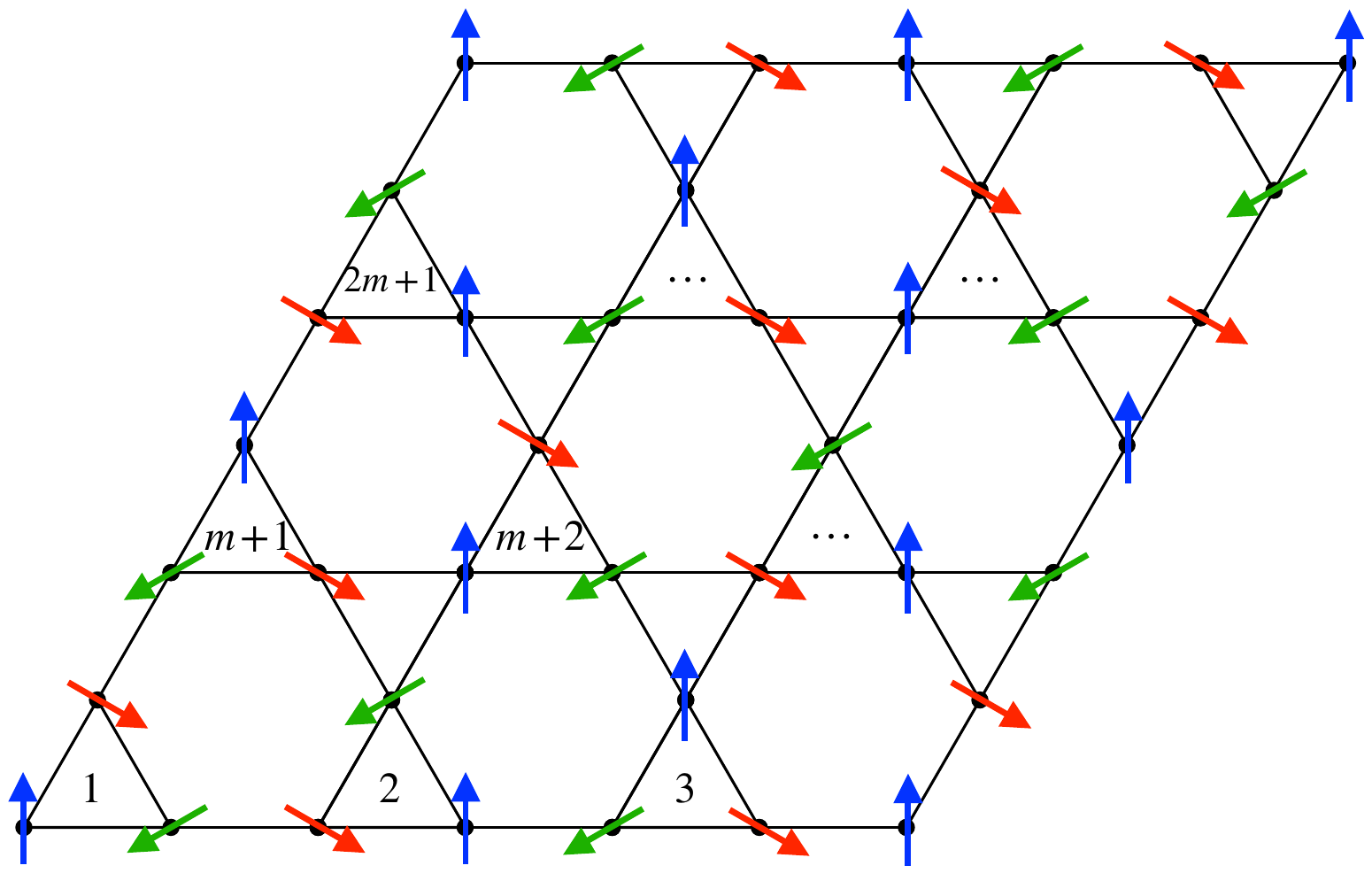}
  \caption{Illustration of a cluster consisting of $m\times m$ kagome unit cells. Numbers on the triangle plaquettes enumerate the unit cells from $1$ to $m^2$.
  A lattice with linear size $L$ is then divided into $\big(\frac{L}{m}\big)^2$ clusters.
  The three-spin cluster used to represent the quadrupolar and octupolar orders in Eqs.~\ref{eq:op_quad} and~\ref{eq:op_oct_c3} corresponds to the special case $m = 1$.
  A perfect $\sqrt{3} \times \sqrt{3}$ state is shown as an illustration, where, for example, the spin pattern in plaquette-$1$ is repeated identically in plaquette-$(m+2)$. }
  \label{fig:mcell}
\end{figure}

\begin{figure}
  \centering
  \includegraphics[width=0.45\textwidth]{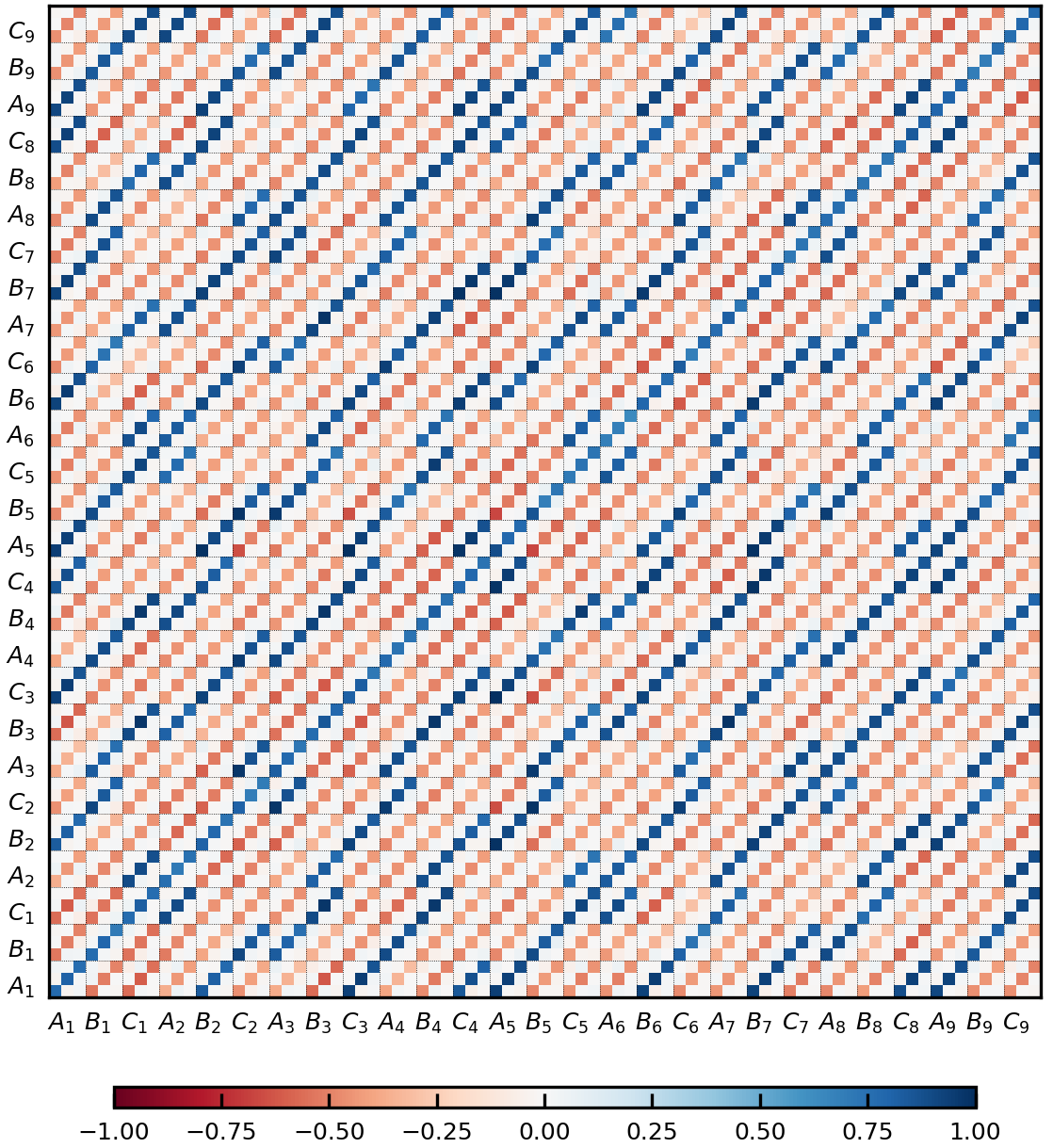}
  \caption{The coefficient matrix $C_{\mu\nu}(T_{\rm low}\,|\,T_\infty)$ learned from a \mbox{rank-1} TK-SVM with a cluster of $3\times 3$ kagome unit cells, where $T_{\rm low} = 10^{-5}J$.
  $A, B, C$ distinguish the three spins (sub-lattices) in a triangular plaquette, enumerating anti-clockwise from the left corner.
  The numbers label unit cells in the cluster, as illustrated in Figure~\ref{fig:mcell}; here $m=3$.
  The $C_{\mu\nu}$ matrix can be divided into $3\times 3$ blocks; each detects the form of correlations between two spins.
  The pattern here reflects a dominant $\sqrt{3} \times \sqrt{3}$ structure in magnetic correlations.
  For example, in the bottom row, the $A_1$-$A_1$ block is positive and has the same structure as the $A_1$-$B_2$ block, indicating that $A_1$ and $B_2$ spin are also positively correlated, as in a $\sqrt{3} \times \sqrt{3}$ state; see Figure~\ref{fig:mcell}.
  Consistently, the $A_5, B_5, C_5$ blocks (plaquette-$(m+2)$) repeat the structure of $A_1, B_1, C_1$ blocks (plaquette-$1$).}
  \label{fig:3cell}
  \clearpage
\end{figure}

\begin{figure}[t]
  \centering
  \includegraphics[width=0.45\textwidth]{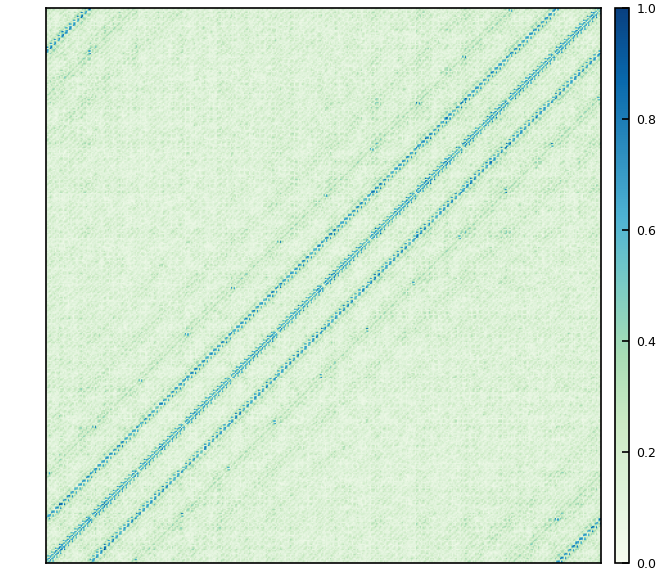}
  \caption{Block structure of the coefficient matrix $C_{\mu\nu}(T_{\rm low}\,|\,T_\infty)$ learned using a \mbox{rank-1} TK-SVM and cluster of $12\times 12$ kagome unit cells.
  Each matrix element is related to the correlation between two spins in the cluster.
  Spins are enumerated from $1$ to $3m^2$ from left (bottom) to right (top), as in Figure~\ref{fig:3cell}; here $m=12$.
  The color (value) of a pixel is defined as the Frobenius norm of the corresponding $3 \times 3$ small block in the full $C_{\mu\nu}(T_{\rm low}\,|\,T_\infty)$.
  }
  \label{fig:12cell}
\end{figure}

\section{Magnetic correlations} \label{sec:sqrt3}
We now examine the magnetic correlations in the low-temperature regime, $T = 10^{-5} J$, of the classical KHAFM, learned by rank-$1$ TK-SVM. 

We first consider a cluster containing $3 \times 3$ kagome unit cells, i.e. $27$ spins (see Figure~\ref{fig:mcell}),  for a system defined on a lattice of linear size $L = 36$.
The resulting $C_{\mu\nu}$ matrix is shown in Figure~\ref{fig:3cell}. It has dimension $81 \times 81$ and can be divided into small $3 \times 3$ blocks.
The structure of those blocks evinces in which way the correlation between two spins is defined.
Because the small $3 \times 3$ blocks have only entries on the diagonal, the correlation is simply captured by the usual inner product $S_a S_b \delta_{ab}$. (In general, the contraction can be more complicated even for magnetic orders).
The change of color between the blocks indicates an antiferromagnetic arrangement of the spins: The blue blocks correspond to positive correlations, and their locations precisely reflect a $\sqrt{3} \times \sqrt{3}$ structure.

Next, in Figure~\ref{fig:12cell}, we show the result using a cluster of $12 \times 12$ kagome unit cells ($432$ spins).
Owing to the large dimensionality of the $C_{\mu\nu}$ matrix, $\dim C_{\mu\nu} = 1296^2$, only the block structure is plotted, where each of its pixels corresponds to a $3 \times 3$ block in Figure~\ref{fig:3cell}.
In the presence of long-range order for an order parameter that can be defined on a small cluster,  $C_{\mu\nu}$ learnt for a large cluster should display the same structure  as for that small cluster.
However, the pattern of Figure~\ref{fig:12cell} does not repeat that of Figure~\ref{fig:3cell}: It fades rather fast, indicating that the linear correlation between two spins is not robustly established at longer distances.
(Note that $C_{\mu\nu}$ does not directly measure the strength of spin-spin correlations. Instead, it probes the form of the correlations and order parameters. See Eqs.~\ref{eq:C_c1r2} and~\ref{eq:Dinfh_Q} for a concrete example.)

Therefore, consistent with observations in the literature~\cite{Huse92, Reimers93, Zhitomirsky08, Chern13, Schnabel12}, our machine detects that the  antiferromagnetic $\sqrt{3} \times \sqrt{3}$ structure is the dominant type of correlation in the dipolar channel  of the classical KHAFM at low temperature.  
However, to establish the  $T \rightarrow 0$ critical point, a systematic analysis of temperature and finite-size effect is needed~\cite{Chern13}, which is beyond the scope of the present paper and current algorithms.

\section{Conclusion} \label{sec:conclusion}

In summary, we have shown how TK-SVM can infer the phase diagram of the classical Heisenberg model defined on the kagome lattice in an unsupervised way. 
It has successfully learned that the spins are constrained to coplanar states owing to an emergent GSC and identified the hidden octupolar and quadrupolar tensor order parameters, which define a $D_{3h}$ biaxial-nematic state.
Moreover, the machine also recognized the dominant $\sqrt{3} \times \sqrt{3}$ correlations in the low temperature regime.

As a data-driven approach, it does not require any prior knowledge or particular insight in the system.
Instead, by revealing the phase diagram and analytical order parameters in an unsupervised setting, the underlying physics becomes immediately evident.
We expect TK-SVM to become an indispensable tool in analyzing many-body spin systems, as it provides an enormous speedup when the order parameters and phase diagrams are complicated.
\ack
We wish to thank Roderich Moessner and Ludovic Jaubert for helpful discussions.
JG, KL, and LP acknowledge support from FP7/ERC Consolidator Grant QSIMCORR, No. 771891, and the Deutsche Forschungsgemeinschaft (DFG, German Research Foundation) under Germany's Excellence Strategy -- EXC-2111 -- 390814868.
Our simulations make use of the $\nu$-SVM formulation~\cite{Scholkopf00}, the LIBSVM library~\cite{Chang01, Chang11}, and the ALPS\-Core library~\cite{Gaenko17}.

\section*{Open source}
The TK-SVM library has been made openly available with documentation and examples~\cite{Jonas}.

\section*{Reference}

\bibliographystyle{iopart-num}
\bibliography{svm-kagome}

\providecommand{\newblock}{}
\begin{thebibliography}{10}
\expandafter\ifx\csname url\endcsname\relax
  \def\url#1{{\tt #1}}\fi
\expandafter\ifx\csname urlprefix\endcsname\relax\def\urlprefix{URL }\fi
\providecommand{\eprint}[2][]{\url{#2}}

\bibitem{BookLacroix}
Lacroix C, Mendels P and Mila F (eds) 2011 {\em Introduction to Frustrated
  Magnetism\/} (Springer Berlin Heidelberg)
  \urlprefix\url{https://doi.org/10.1007/978-3-642-10589-0}

\bibitem{Balents10}
Balents L 2010 {\em Nature\/} {\bf 464} 199--208

\bibitem{Chalker92}
Chalker J~T, Holdsworth P~C~W and Shender E~F 1992 {\em Phys. Rev. Lett.\/}
  {\bf 68}(6) 855--858
  \urlprefix\url{https://link.aps.org/doi/10.1103/PhysRevLett.68.855}

\bibitem{Villain79}
Villain J 1979 {\em Z. Phys. B\/} {\bf 33} 31--42 ISSN 1431-584X

\bibitem{Villain80}
Villain J, Bidaux R, Carton J~P and Conte R 1980 {\em J. Phys. France\/} {\bf
  41} 1263--1272

\bibitem{Reimers93}
Reimers J~N and Berlinsky A~J 1993 {\em Phys. Rev. B\/} {\bf 48}(13) 9539--9554
  \urlprefix\url{https://link.aps.org/doi/10.1103/PhysRevB.48.9539}

\bibitem{Huse92}
Huse D~A and Rutenberg A~D 1992 {\em Phys. Rev. B\/} {\bf 45}(13) 7536--7539
  \urlprefix\url{https://link.aps.org/doi/10.1103/PhysRevB.45.7536}

\bibitem{Ritchey93}
Ritchey I, Chandra P and Coleman P 1993 {\em Phys. Rev. B\/} {\bf 47}(22)
  15342--15345
  \urlprefix\url{https://link.aps.org/doi/10.1103/PhysRevB.47.15342}

\bibitem{Berezinskii71}
{Berezinski{\v{i}}} V~L 1971 {\em Soviet Journal of Experimental and
  Theoretical Physics\/} {\bf 32} 493

\bibitem{Berezinskii72}
{Berezinski{\v{i}}} V~L 1972 {\em Soviet Journal of Experimental and
  Theoretical Physics\/} {\bf 34} 610

\bibitem{Kosterlitz73}
Kosterlitz J~M and Thouless D~J 1973 {\em Journal of Physics C: Solid State
  Physics\/} {\bf 6} 1181--1203
  \urlprefix\url{https://doi.org/10.1088%2F0022-3719%2F6%2F7%2F010}

\bibitem{Mermin79}
Mermin N~D 1979 {\em Rev. Mod. Phys.\/} {\bf 51}(3) 591--648
  \urlprefix\url{http://link.aps.org/doi/10.1103/RevModPhys.51.591}

\bibitem{Michel80}
Michel L 1980 {\em Rev. Mod. Phys.\/} {\bf 52}(3) 617--651
  \urlprefix\url{https://link.aps.org/doi/10.1103/RevModPhys.52.617}

\bibitem{Zhitomirsky02}
Zhitomirsky M~E 2002 {\em Phys. Rev. Lett.\/} {\bf 88}(5) 057204
  \urlprefix\url{https://link.aps.org/doi/10.1103/PhysRevLett.88.057204}

\bibitem{Zhitomirsky08}
Zhitomirsky M~E 2008 {\em Phys. Rev. B\/} {\bf 78}(9) 094423
  \urlprefix\url{https://link.aps.org/doi/10.1103/PhysRevB.78.094423}

\bibitem{Henley10}
Henley C~L 2010 {\em Annu. Rev. Condens. Matter Phys.\/} {\bf 1} 179--210

\bibitem{Castelnovo12}
Castelnovo C, Moessner R and Sondhi S~L 2012 {\em Annu. Rev. Condens. Matter
  Phys.\/} {\bf 3} 35--55

\bibitem{Henley09}
Henley C~L 2009 {\em Phys. Rev. B\/} {\bf 80}(18) 180401
  \urlprefix\url{https://link.aps.org/doi/10.1103/PhysRevB.80.180401}

\bibitem{Chern13}
Chern G~W and Moessner R 2013 {\em Phys. Rev. Lett.\/} {\bf 110}(7) 077201
  \urlprefix\url{https://link.aps.org/doi/10.1103/PhysRevLett.110.077201}

\bibitem{Schnabel12}
Schnabel S and Landau D~P 2012 {\em Phys. Rev. B\/} {\bf 86}(1) 014413
  \urlprefix\url{https://link.aps.org/doi/10.1103/PhysRevB.86.014413}

\bibitem{Ponte17}
Ponte P and Melko R~G 2017 {\em Phys. Rev. B\/} {\bf 96}(20) 205146
  \urlprefix\url{https://link.aps.org/doi/10.1103/PhysRevB.96.205146}

\bibitem{Wang16}
Wang L 2016 {\em Phys. Rev. B\/} {\bf 94}(19) 195105
  \urlprefix\url{https://link.aps.org/doi/10.1103/PhysRevB.94.195105}

\bibitem{Nieuwenburg17}
van Nieuwenburg E~P~L, Liu Y~H and Huber S~D 2017 {\em Nat. Phys.\/} {\bf 13}
  435--439

\bibitem{Carrasquilla17}
Carrasquilla J and Melko R~G 2017 {\em Nat. Phys.\/} {\bf 13} 431--434

\bibitem{LiuYH18}
Liu Y~H and van Nieuwenburg E~P~L 2018 {\em Phys. Rev. Lett.\/} {\bf 120}(17)
  176401
  \urlprefix\url{https://link.aps.org/doi/10.1103/PhysRevLett.120.176401}

\bibitem{Rodriguez-Nieva19}
Rodriguez-Nieva J~F and Scheurer M~S 2019 {\em Nature Physics\/} {\bf 15}
  790--795 \urlprefix\url{https://doi.org/10.1038/s41567-019-0512-x}

\bibitem{Zhang20}
Zhang Y, Ginsparg P and Kim E~A 2020 {\em Phys. Rev. Research\/} {\bf 2}(2)
  023283
  \urlprefix\url{https://link.aps.org/doi/10.1103/PhysRevResearch.2.023283}

\bibitem{CarleoTroyer17}
Carleo G and Troyer M 2017 {\em Science\/} {\bf 355} 602--606 ISSN 0036-8075
  (\textit{Preprint}
  \eprint{http://science.sciencemag.org/content/355/6325/602})
  \urlprefix\url{http://science.sciencemag.org/content/355/6325/602}

\bibitem{Huang17}
Huang L and Wang L 2017 {\em Phys. Rev. B\/} {\bf 95}(3) 035105
  \urlprefix\url{https://link.aps.org/doi/10.1103/PhysRevB.95.035105}

\bibitem{Cai18}
Cai Z and Liu J 2018 {\em Phys. Rev. B\/} {\bf 97}(3) 035116
  \urlprefix\url{https://link.aps.org/doi/10.1103/PhysRevB.97.035116}

\bibitem{Melko19}
Melko R~G, Carleo G, Carrasquilla J and Cirac J~I 2019 {\em Nature Physics\/}
  {\bf 15} 887--892 \urlprefix\url{https://doi.org/10.1038/s41567-019-0545-1}

\bibitem{Pfau19}
Pfau D, Spencer J~S, de~G~Matthews A~G and Foulkes W~M~C 2019 Ab-initio
  solution of the many-electron schrödinger equation with deep neural networks
  (\textit{Preprint} \eprint{1909.02487})

\bibitem{Hermann19}
Hermann J, Schätzle Z and Noé F 2019 Deep neural network solution of the
  electronic schrödinger equation (\textit{Preprint} \eprint{1909.08423})

\bibitem{Liao19}
Liao H~J, Liu J~G, Wang L and Xiang T 2019 {\em Phys. Rev. X\/} {\bf 9}(3)
  031041 \urlprefix\url{https://link.aps.org/doi/10.1103/PhysRevX.9.031041}

\bibitem{CarleoChoo19}
Carleo G, Choo K, Hofmann D, Smith J~E, Westerhout T, Alet F, Davis E~J,
  Efthymiou S, Glasser I, Lin S~H, Mauri M, Mazzola G, Mendl C~B, {van
  Nieuwenburg} E, O’Reilly O, Théveniaut H, Torlai G, Vicentini F and Wietek
  A 2019 {\em SoftwareX\/} {\bf 10} 100311 ISSN 2352-7110
  \urlprefix\url{http://www.sciencedirect.com/science/article/pii/S2352711019300974}

\bibitem{Nagai17}
Nagai Y, Shen H, Qi Y, Liu J and Fu L 2017 {\em Phys. Rev. B\/} {\bf 96}(16)
  161102 \urlprefix\url{https://link.aps.org/doi/10.1103/PhysRevB.96.161102}

\bibitem{Xu17}
Xu X~Y, Qi Y, Liu J, Fu L and Meng Z~Y 2017 {\em Phys. Rev. B\/} {\bf 96}(4)
  041119 \urlprefix\url{https://link.aps.org/doi/10.1103/PhysRevB.96.041119}

\bibitem{Xie18}
Xie T and Grossman J~C 2018 {\em Phys. Rev. Lett.\/} {\bf 120}(14) 145301
  \urlprefix\url{https://link.aps.org/doi/10.1103/PhysRevLett.120.145301}

\bibitem{Lee16}
Lee J, Seko A, Shitara K, Nakayama K and Tanaka I 2016 {\em Phys. Rev. B\/}
  {\bf 93}(11) 115104
  \urlprefix\url{https://link.aps.org/doi/10.1103/PhysRevB.93.115104}

\bibitem{Isayev17}
Isayev O, Oses C, Toher C, Gossett E, Curtarolo S and Tropsha A 2017 {\em
  Nature Communications\/} {\bf 8} 15679
  \urlprefix\url{https://doi.org/10.1038/ncomms15679}

\bibitem{Zhu18}
Zhu Q, Samanta A, Li B, Rudd R~E and Frolov T 2018 {\em Nature
  Communications\/} {\bf 9} 467
  \urlprefix\url{https://doi.org/10.1038/s41467-018-02937-2}

\bibitem{Carleo19}
Carleo G, Cirac I, Cranmer K, Daudet L, Schuld M, Tishby N, Vogt-Maranto L and
  Zdeborov\'a L 2019 {\em Rev. Mod. Phys.\/} {\bf 91}(4) 045002
  \urlprefix\url{https://link.aps.org/doi/10.1103/RevModPhys.91.045002}

\bibitem{Schmidt19}
Schmidt J, Marques M~R~G, Botti S and Marques M~A~L 2019 {\em npj Computational
  Materials\/} {\bf 5} 83
  \urlprefix\url{https://doi.org/10.1038/s41524-019-0221-0}

\bibitem{Carrasquilla20}
Carrasquilla J 2020 {\em Advances in Physics: X\/} {\bf 5} 1797528
  \urlprefix\url{https://doi.org/10.1080/23746149.2020.1797528}

\bibitem{Greitemann19}
Greitemann J, Liu K and Pollet L 2019 {\em Phys. Rev. B\/} {\bf 99}(6)
  060404(R) \urlprefix\url{https://link.aps.org/doi/10.1103/PhysRevB.99.060404}

\bibitem{Liu19}
Liu K, Greitemann J and Pollet L 2019 {\em Phys. Rev. B\/} {\bf 99}(10) 104410
  \urlprefix\url{https://link.aps.org/doi/10.1103/PhysRevB.99.104410}

\bibitem{Greitemann19b}
Greitemann J, Liu K, Jaubert L~D~C, Yan H, Shannon N and Pollet L 2019 {\em
  Phys. Rev. B\/} {\bf 100}(17) 174408
  \urlprefix\url{https://link.aps.org/doi/10.1103/PhysRevB.100.174408}

\bibitem{Liu20}
Liu K, Sadoune N, Rao N, Greitemann J and Pollet L 2020 {\em arXiv preprint
  arXiv:2004.14415\/}

\bibitem{Jonas_thesis}
Greitemann J 2019 Investigation of hidden multipolar spin order in frustrated
  magnets using interpretable machine learning techniques
  \urlprefix\url{http://nbn-resolving.de/urn:nbn:de:bvb:19-250579}

\bibitem{CortesVapnik95}
Cortes C and Vapnik V 1995 {\em Machine Learning\/} {\bf 20} 273--297 ISSN
  1573-0565 \urlprefix\url{https://doi.org/10.1007/BF00994018}

\bibitem{BookVapnik}
Vapnik V~N 1998 {\em Statistical learning theory\/} (Wiley) ISBN 9780471030034
  \urlprefix\url{https://www.wiley-vch.de/en?option=com_eshop&view=product&isbn=9780471030034&title=Statistical%20Learning%20Theory}

\bibitem{Liu16}
Liu K, Nissinen J, Slager R~J, Wu K and Zaanen J 2016 {\em Phys. Rev. X\/} {\bf
  6}(4) 041025
  \urlprefix\url{https://link.aps.org/doi/10.1103/PhysRevX.6.041025}

\bibitem{Nissinen16}
Nissinen J, Liu K, Slager R~J, Wu K and Zaanen J 2016 {\em Phys. Rev. E\/} {\bf
  94}(2) 022701
  \urlprefix\url{http://link.aps.org/doi/10.1103/PhysRevE.94.022701}

\bibitem{Michel01}
Michel L 2001 {\em Phys. Rep.\/} {\bf 341} 11--84

\bibitem{Bottou07}
Bottou L and Lin C~J 2007 {\em Large scale kernel machines\/} {\bf 3} 301--320

\bibitem{HsuLin02}
Hsu C~W and Lin C~J 2002 {\em IEEE transactions on Neural Networks\/} {\bf 13}
  415--425

\bibitem{Fiedler73}
Fiedler M 1973 {\em Czechoslovak Mathematical Journal\/} {\bf 23} 298--305
  \urlprefix\url{http://eudml.org/doc/12723}

\bibitem{Fiedler75}
Fiedler M 1975 {\em Czechoslovak Mathematical Journal\/} {\bf 25} 619--633
  \urlprefix\url{http://eudml.org/doc/12900}

\bibitem{BookDeGennes}
de~Gennes P and Prost J 1995 {\em The Physics of Liquid Crystals\/}
  International Series of Monographs on Physics (Clarendon Press) ISBN
  9780198517856

\bibitem{Fel95}
Fel L~G 1995 {\em Phys. Rev. E\/} {\bf 52}(1) 702--717
  \urlprefix\url{http://link.aps.org/doi/10.1103/PhysRevE.52.702}

\bibitem{Scholkopf00}
Sch{\"o}lkopf B, Smola A~J, Williamson R~C and Bartlett P~L 2000 {\em Neural
  Comput.\/} {\bf 12} 1207--1245

\bibitem{Chang01}
Chang C~C and Lin C~J 2001 {\em Neural Comput.\/} {\bf 13} 2119--2147

\bibitem{Chang11}
Chang C~C and Lin C~J 2011 {\em ACM Trans. Intell. Syst. Technol.\/} {\bf 2}
  27:1--27:27 ISSN 2157-6904
  \urlprefix\url{http://doi.acm.org/10.1145/1961189.1961199}

\bibitem{Gaenko17}
Gaenko A, Antipov A, Carcassi G, Chen T, Chen X, Dong Q, Gamper L, Gukelberger
  J, Igarashi R, Iskakov S, K{\"o}nz M, LeBlanc J, Levy R, Ma P, Paki J,
  Shinaoka H, Todo S, Troyer M and Gull E 2017 {\em Comput. Phys. Commun.\/}
  {\bf 213} 235--251 ISSN 0010-4655
  \urlprefix\url{http://www.sciencedirect.com/science/article/pii/S0010465516303885}

\bibitem{Jonas}
Greitemann J, Liu K and Pollet L tensorial-kernel SVM library,
  \url{https://gitlab.physik.uni-muenchen.de/tk-svm/tksvm-op}

\end{thebibliography}
\end{document}